\documentclass{aa} 
\usepackage{natbib}
\usepackage{xcolor}
\usepackage{graphicx}
\usepackage{txfonts}
\usepackage[]{hyperref}
\begin{document}

   \title{Radial velocities: direct application of Pierre Connes' shift finding algorithm to Cross-Correlation Functions}

   \author{Jean-Loup Bertaux
          \inst{1,2}
          \and
          Anastasiia Ivanova
          \inst{2,3} 
          \and
          Rosine Lallement 
          \inst{4}
          }

   \institute{LATMOS, Sorbonne Universit\'e, 4 Place Jussieu, 75005 Paris, France\\
              \email{jean-loup.bertaux@latmos.ipsl.fr}
         \and
             LATMOS, Universit\'e Versailles-Saint-Quentin, 11 Bd D'Alembert, 78280 Guyancourt, France\\
        \and Space Research Institute (IKI), Russian Academy of Science, Moscow 117997, Russia
        \email{anastasia.ivanova@cosmos.ru}
        \and   GEPI, Observatoire de Paris, PSL University, CNRS,  5 Place Jules Janssen, 92190 Meudon, France
             }

   \date{Received , ; accepted }
\titlerunning{A new method for determining stellar radial velocity changes}

 
\abstract
{Pipelines of state-of-the-art spectrographs dedicated to planet detection provide, for each exposure, series of Cross-Correlation Functions (CCFs) built with a Binary Mask (BM), as well as the absolute radial velocity (RV) derived from Gaussian fit of a weighted average CCF$_{tot}$ of the CCFs.}
{Our aim was to test the benefits of the application of the shift finding algorithm developed by Pierre Connes directly to the total CCF$_{tot}$, and to compare the resulting RV shifts (DRVs) with the results of the Gaussian fits. In a second step, we investigated how the individual DRV profiles along the velocity grid derived from the shift finding algorithm can be used as an easy tool for detection of stellar line shape variations.}
{We developed the corresponding algorithm and tested it on 1151 archived spectra of the K2.5 V star HD 40307 obtained with ESO/ESPRESSO during a one-week campaign in 2018. Tests were performed based on the comparison of DRVs with RVs from Gaussian fits. DRV profiles along the velocity grid (DRV(i)) were scrutinized and compared with direct CCF$_{tot}$ ratios.}
{The dispersion of residuals from a linear fit to RVs from 406 spectra recorded within a single night, a measure of mean error, was found to be $\sigma$=1.03 and 0.83 ms$^{-1}$ for the Gaussian fit and the new algorithm respectively, a significant 20\% improvement in accuracy. The two full one-week series obtained during the campaign were also fitted with a 3-planet system Keplerian model. The residual divergence between data and best-fit model is significantly smaller for the new algorithm than for the Gaussian fit. Such a difference was found to be associated in a large part with an increase by $\simeq$ 1.3 m.s$^{-1}$ of the difference between the two types of RV values between the third and fourth nights. Interestingly, the DRV(i) profiles reveal at the same time a significant variation of line shape.}
{ The shift finding algorithm is a fast and easy tool allowing to obtain additional diagnostics on the RV measurements in series of exposures. For observations made in the same instrumental configuration and if line shapes are not varying significantly, it increases the accuracy of velocity variation determinations. On the other hand, departures from constancy of the DRVi profiles, as well as varying differences between RVs from this new method and RVs from a Gaussian fit can detect and report in a simple way line shape variations due to stellar activity. }

\maketitle

\section{Introduction}
Since the discovery of 51 Pegasi b, the first exoplanet detected around a main-sequence star \citep{Mayor1995}, many other planets (1065, 22 September 2023) have been detected with the indirect method of monitoring the radial velocity RV (which monitors the reflex motion of the star), and their projected mass determined (m $\sin$ i). Since the reflex RV is proportional to the planet mass m, there is a great interest to increase the precision of RV measurements, in order to approach the case of an Earth’s twin planet in the habitable zone, about 10 cms$^{-1}$.
Modern spectrographs, like ESPRESSO at the Very Large Telescope \citep{Pepe2021} or HARPS North and South \citep{Pepe2000,Cosentino12}, have reached unprecedented stability and precision in wavelength assignment of observed spectra, well below 1 ms$^{-1}$. Another way to increase the precision is to add new wavelength domains to the optical domain. This is the case of new near IR (NIR) spectrographs like SPIROU \citep{Donati2020} and NIRPS \citep{Bouchy2017}). Correction of telluric absorptions, in replacement of rejection of contaminated spectral regions, is now also increasing the RV accuracy, moderately in the visible (see, e.g., \citet{Ivanova2023} for ESPRESSO), and strongly in the NIR (see \citet{Cook22} for SPIROU).\\
However, now that the spectrometers instruments have reached a precision well below 1 ms$^{-1}$, it appears that the time variations of the blueshifts associated with convective granulation and supergranulation of the star atmosphere are the limiting factors for the detection of small mass planets by means of RV methods \citep{Meunier15}, because they affect the shifts of the stellar lines from which we attempt to derive a change of Doppler shift of spectral lines as a proxy of the dynamical dR/dt. The granulation is the irregular cellular pattern at the surface of stars that arise because stars have a convective envelope in their photospheres where the hotter bubbles of gas rise (blueshift) and the cooler bubbles sink (redshift). The imbalance between contribution of hot granules and cool intergranular lanes leads to the blueshift of RV or convective granulation blueshift, GBS in the following \citep[see][]{Dravins81}. The blue shift depends on the particular line, may amount up to 400-600 ms$^{-1}$, and is correlated with the Equivalent Width (EW) of the line: large EW lines have almost no blue shift, small EW lines have the largest blue shifts, as measured on the integrated sun \citep{Gonzalez2020}.\\
Two main processes affecting the RV measurements connected to GBS may be identified, acting with different time scales:
1. Stochastic fluctuations of granulation and supergranulation convective blueshift. The life time of one granule is of the order of a few minutes, and the number of granules on the visible disc of a star is on the order of 10$^{4}$-10$^{5}$ for a star like the Sun. The average RV will not be rigorously constant, but rather will display stochastic fluctuations. The magnitude of these fluctuations has been estimated from model simulations by \cite{Meunier2020} for stars from F6 to K4 spectral classes: peak-to-peak fluctuations of the order of 2 to 0.5 ms$^{-1}$, respectively. 
2. Attenuation of granulation blue shift.
It has been shown from solar observations that the convective granulation blueshift is inhibited or attenuated by magnetic activity, which manifests itself on the Sun by solar spots, faculae and plages \citep{Meunier2021}. From analysis of solar Dopplergrams and magnetograms obtained with the MDI instrument of board of SOHO \cite{Meunier10} have measured along a whole solar cycle (11 years) that the average RV of the integrated solar disc is increasing with solar activity (due to the attenuation of the granulation blueshift) from 0 (quiet Sun as a reference) up to 8 ms$^{-1}$, almost 100 times larger than the reflex motion induced by Earth around a G star. Therefore, as said above, the granulation and supergranulation and their time variations are the limiting factors for the detection of Earth size planets via RV methods \citep{Meunier15}.\\
Coming back to the pioneering work of \cite{Dravins81} studying high resolution solar spectra, they argued that “the bottoms of strong absorption lines should show smaller shifts since they form high up in the atmosphere where the granulation is no longer distinctly visible” and “the lines in different wavelength regions with different granular/intergranular contrast or different atmospheric opacity should show different amounts of line asymmetries and shifts”. Therefore, they explained physically why the solar lines are distorted, as evidenced by their bisector showing a characteristic C shape. At that time, however, they did not discussed changes of these distortions, but pretty soon exo-planet hunters used the bisector analysis to attempt to discriminate dynamical RV changes due to the presence of exo-planets from spurious, activity connected, RV changes \citep[e.g.][]{Queloz01}. \\
One of the methods to analyze a RV time series of observations of a particular star is based on the construction of a binary mask (BM) around selected stellar lines and the computation of the cross-correlation function between mask windows and the observed stellar lines. Gaussian fitting of the weighted average of the various CCFs (herafter CCF$_{tot}$) provides the absolute radial velocity RVabs. At present, official pipelines of the most sophisticated spectrographs (for example, the ESPRESSO pipeline\footnote{https://www.eso.org/sci/facilities/paranal/instruments/espresso/doc.html}) make available the results of the Gaussian fit of the CCF$_{tot}$, and additionally the CCFs for each order and their weighted sum CCF$_{tot}$. Users can directly use the pipeline RVs for planetary orbit modeling. These pipeline values provide also on-the-fly quality checks of the measurements.\\
There are other techniques of precise RV (pRV) measurements than the BM/CCF method, some of them reaching a high degree of complexity and outperforming the use of CCFs. Template fitting is also largely developed and used in pipelines \citep{Zechmeister18, Astudillo15}. Data-driven techniques \citep{Bedell19}  and line-by-line (LBL) analyses initiated by \cite{Dumusque2018} and developed further by \cite{Artigau22}, as well as the YARARA approach of \citet{Cretignier21}, are other sophisticated post-processing methods reaching very high accuracy. We do not detail the different methods here, except for the line-by-line algorithm  because there is one common point with our work, namely the use of the shift finding algorithm developed by Pierre Connes \citep{Connes1985}. In the field of precise RV measurements, \citet{Artigau22} have developed a method which is very efficient at eliminating the outliers, whatever the cause of them. First is built a high signal-to-noise ratio (SNR) template spectrum of the star, by taking the median of all observed spectra. Each stellar spectrum is BERV-registered. Then, each current spectrum is compared, line-by-line, to the template, and one RV value (actually, a change of RV w.r.t. the line template) is derived for each line with the Connes shift finding algorithm, revisited by \cite{Bouchy2001}. A histogram of the DRV values for the current spectrum is built, and a sigma-clipping is done, eliminating all DRV values defined as outliers. Outliers may come from telluric residuals, cosmic rays, detector defects, and other features. The stellar activity, responsible for spurious shifts, is mitigated by using a Gaussian Process fitted to the time series of an activity indicator, linked to the variation of the average FWHM of all lines (in the velocity scale).
The CCF method keeps some advantages. Compared to individual lines, merged CCFs have very high signal and SNR. Moreover, the RV extraction does not need auxiliary data and no post-processing using such data. As such, it is appropriate for on-the-fly measurements or analysis of limited measurements of the same target, and for pipeline products. Finally, full CCFs can be used to investigate the line shape and its variability, in addition to the simple Gaussian fits.\\
The idea that we explore here is to apply the shift finding algorithm to the CCF$_{tot}$ (and not to individual lines, at variance with \cite{Artigau22}). It can be seen as an algorithm of intermediate complexity between the Gaussian fit of pipeline-based CCFs and the LBL  method. It combines the advantage of the CCF technique, in particular the high signal, and the optimal extraction of RV change from the comparison between two spectra. It should be rather simple for the users to explore the capabilities of this technique with the corresponding Python software available along with this article, using either archived data or their own observations.
We do not pretend that it would give more precise RV values than the most sophisticated methods mentioned above. Our aim is to show that it can do better than the CCF Gaussian fit, and bring some on-the-fly warning information on spectral line shape variations. The overall objective of this paper is to describe this algorithm, and explore its capabilities on a particular example of a RV time series. In homage to the pioneering works of Pierre Connes in the field of exo-planets and other instrumentation for astronomy, e.g. Fourier transform spectroscopy, we named this algorithm EPiCA (Exoplanets Pierre Connes’ Algorithm).\\
In Section \ref{sec:method} we recall the shift finding approach and remind the two basic formulas of his work: the first formula is retrieving the change of RV between two spectra, and the second formula gives an intrinsic maximum precision provided by a piece of spectrum and the number of photons collected. In Section \ref{sec:ccf} we remind the classical RV method, using the change of minima of Gaussian fit to CCF. It is the one used in the official ESO pipeline. In Section \ref{sec:proposed} we describe our algorithm, and validate it by a simulation exercise. In Section \ref{sec:application}, our method is applied to a time series of RV measurements on star HD 40307, known to host planets. The achieved accuracy on the RV variation is tested in two ways, by comparing the RV dispersion with the one resulting from a Gaussian fit on the one hand, and by comparing the best-fit to the three-planets Keplerian model based on RVs from the algorithm and the one based on RVs from a Gaussian fit. In section \ref{variability} we describe how the method allows an easy detection of line shape variability. A large part of the data analysis of this paper is derived from Ivanova (2023) PhD manuscript \footnote{\href{PHD manuscript}{https://theses.hal.science/tel-04504477}}.\\
We provide a Python code implementing this new algorithm, freely accessible through the GitHub repository. \footnote{\href{EPiCA website}{https://github.com/aeictf/EPiCA}}.

\section{The shift finding method to measure star Radial Velocity changes.}
\label{sec:method}
\begin{figure}
\includegraphics[width=0.5\textwidth]{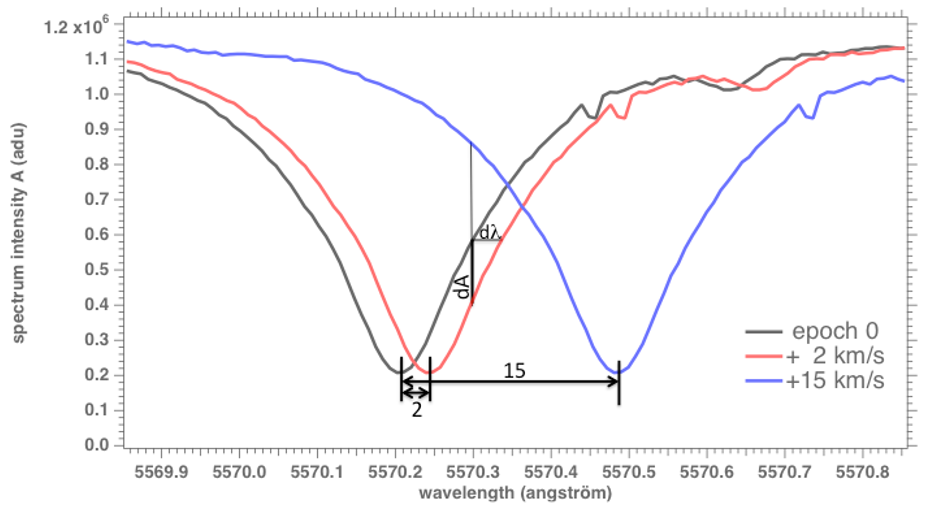}
\caption{Example of a spectral line at 3 epochs. Epoch 0 - reference (black), Epoch 1 - shifted by +2 kms$^{-1}$ (red), Epoch 2 - shifted by +15 kms$^{-1}$ (blue). The intensity change is measured within the d$\lambda$ slice. This figure is similar to the figure 4 from \citet{Connes1985} }
\label{fig:fig1}
\end{figure}

This method is fully described in a seminal paper \citep{Connes1985}. At that time no exoplanet had been discovered, in spite of various attempts. In this paper the author said that the best way to discover an exoplanet would be an indirect method: monitoring for periodic variations of the radial velocity RV of a star. He was even more specific, saying that this could be achieved by measuring the spectrum of the star with a high-resolution spectrometer, with an echelle-grating spectrograph, cross-dispersed, and a CCD detector.\\
He also said that it may be illusory to measure the absolute RV of a star, dR/dt, (R, distance of the star to the observer) with an accuracy of 1 ms$^{-1}$ from the observation of a star spectrum, because it is only a proxy of the mechanical dR/dt. For instance, clearly the Einstein effect cannot be estimated accurately enough to predict its magnitude for a given star which surface gravity may be known by other means: its spectral type. This general relativity Einstein effect is mimicking exactly a Doppler shift, and cannot be disentangled from a Doppler effect. It is of the order of 600 ms$^{-1}$ for the Sun \citep{Gonzalez2020}.\\
However, the {\it variation} of RV with time may be accurately measured, making the difference of two determinations of RV at two epochs: the Einstein effect, a constant for a given star, disappears from the difference. 
This is why Pierre Connes called his method: "Accelerometry". However, it is a somewhat improper denomination, since acceleration is a $\Delta$RV/$\Delta$t, implying a given $\Delta$t. Indeed, we need to measure accurately a $\Delta$RV=RV$_1$-RV$_2$ during two epochs E$_1$ and E$_2$, whatever the $\Delta$t between the two epochs. We may guess that the basic reason for keeping the word "Accelerometry" is for the beauty of the acronym: AAA, Absolute Astronomical Accelerometer. 
Having said that it was needed to acquire two high resolution spectra of the star on a CCD, at epochs E$_1$ and E$_2$, \citet{Connes1985} designed an algorithm (we will call it the shift finding algorithm) to retrieve the difference of RV of the two spectra. It is based on the variation of intensity in a pixel, (or a spectel at a given wavelength) associated to the slope of the spectrum.\\
This is schematized on figure 4 of \citet{Connes1985}, and on Figure \ref{fig:fig1} here.

If a spectrum is displaced along the wavelength axis, the intensity on one pixel (or spectel) will decrease or increase, as a function of the displacement and of the derivative of the spectrum along the axis. \\
In the following, we will have new notations:

\begin{itemize}
    \item the axis is the wavelength along the 2D array detector lines, with continuous coordinate sp (for spectel)
    \item the spectel index is i
    \item the intensity of the spectrum is A$_0$ at epoch 0, and A$_n$ at another epoch n. This intensity is to be measured in electrons created by light. Both spectra A$_0$ and A$_n$ must be identical, only shifted w.r.t. one another by a small displacement $\delta$sp
\end{itemize}

The intensity in one pixel i (or one spectel) is changing by dA(i) given by:
\begin{equation}
    dA(i)=A_n(i)-A_0(i)=-\frac{\partial A_0(i)}{\partial sp}\delta sp(i)
    \label{eq:deltaa}
\end{equation}
So it possible to extract for each pixel i (or spectel) the wavelength shift $\delta$sp(i). We call it the first formula of Connes, replacing A$_{0}$ by A:
\begin{equation}
    \delta sp(i)=-\frac{A_n(i)-A(i)}{\partial A(i)/\partial sp}
    \label{eq:deltasp}
\end{equation}
Equation \ref{eq:deltasp} can be transformed into a Doppler shift and the corresponding change $\delta$V(i) of the radial velocity of the star between two epochs:
\begin{equation}
    \frac{\delta V_i}{c} = \frac{\delta sp(i)}{sp(i)} = -\frac{A_n(i)-A(i)}{sp(i)\partial A(i)/\partial sp}
    \label{eq:newdeltasp}
\end{equation}

Then all the $\frac{\delta V(i)}{c}$ have to be combined accounting for their individual errors  $\sigma$(i). In the case of Gaussian errors, it is known that the optimal combination is to use weights = 1/ $\sigma^2$(i), while the error $\sigma$(RV) on the combined retrieved RV is such that 
\begin{equation}
    \frac{1}{\sigma^2}=\sum_i \frac{1}{\sigma(i)^2}
\end{equation}
The variance $\sigma$$^{2}$(i) of $\frac{\delta V(i)}{c}$ can be calculated by the relation Var(aX+bY)=a$^2$Var(X)+b$^2$Var(Y) where a and b are constants and X and Y are random variables, applied to Equation \ref{eq:newdeltasp}. 
\begin{equation}
    \begin{aligned}
    \sigma^2(i)=(\frac{1}{sp(i)  \partial A(i) / \partial sp(i)})^2 Var(A(i))\\
    +(\frac{1}{sp(i)  \partial A(i) / \partial sp(i)})^2 Var(A_n(i))
    \end{aligned}
    \label{eq:sigma2}
\end{equation}
The variance of a number of photoelectrons A is equal to A. In most practical cases, the reference spectrum A(i) is built from many spectra and Var(A(i))  becomes much smaller than Var(A$_n$(i)). Hence the equation \ref{eq:sigma2} will be:
\begin{equation}
    \sigma^2(i)= (\frac{1}{sp(i)  \partial A(i) / \partial sp(i)})^2 A_n(i)
    \label{eq:sigma2fin}
\end{equation}
Here, it is assumed that A(i) is a perfect noise-free spectrum. This is always the case when a template median spectrum is build from several tens of observations, as practiced commonly by high-precision RV programs. In the case of the comparison of two spectra of similar intensities (for instance, taking the first of a time series as a reference),  then a factor 2 should be added as a multiplier in equation \ref{eq:sigma2fin}.\\
Following the notations of \citet{Bouchy2001} who revisited the Connes approach, the next step is to introduce the weight function W(i).
\begin{equation}
    W(i)=\frac{1}{\sigma(i)^2}=\frac{(sp(i)  \partial A(i) / \partial sp(i))^2 }{A_n(i)}
\end{equation}
The velocity change then is:
\begin{equation}
    \frac{\delta V}{c}=\frac{\sum \frac{\delta V(i)}{c}W(i)}{\sum W(i)}=
    \frac{\sum (A(i)-A_n(i)) (\frac{W(i)}{A_n})^{1/2}}{\sum W(i)}
\label{equation8}
\end{equation}

\cite{Connes1985} also showed that the uncertainty on a RV measurements can be calculated {\it a priori} on the basis of photon noise. Even if there is no velocity change in between epochs due to the motion, we can find a small velocity change caused by the noise perturbation of spectrum at epoch n. 
In the following we consider only the photon noise, neglecting the readout noise from the detector.
The approach is based on the quality factor Q which can be computed for any star, and is independent from the absolute flux (if we neglect the detector noise contribution).The quality factor Q depends on the wavelength structure of the spectrum and $\delta V_{min}$ is the photon noise limit which can be achieved to determine the absolute wavelength position of a piece of an observed spectrum in radial velocity units:
\begin{equation}
    \delta V_{min}=\frac{c}{Q\sqrt{N_{e^{-}}}}
    \label{eq:secondformula}
\end{equation}
where c is the speed of light,   N$_{e^{-}}$ is the total number of photoelectrons counted over the whole spectral range considered, and Q is the quality factor equal to : 
\begin{equation}
    Q= \frac{\sqrt{\sum W(i)}}{\sqrt{\sum A(i)}}
\end{equation}
We call it the second formula of Connes.
It should be noted that the observed spectra are given in ADU units. In order to convert into a number of electrons, one must multiply by the gain of the CCF read-out system, the number of electrons per ADU, which is included in the headers of ESPRESSO spectra (0.9 electrons per ADU in the present case). 
\begin{equation}
    \delta V_{min}=\frac{c}{Q\sqrt{N_{e^{-}}}}=\frac{c}{Q\sqrt{0.9*N_{ADU}}}
\end{equation}
\citet{Connes1985} demonstrated mathematically that this algorithm is optimal (the best possible), and makes use of all photons and information contained in the spectrum in an optimal way. This fact has been later recognized many times by scientists working on the subject, for instance those who used the Cross-Correlation Function of the spectrum with a spectral line which wavelength is well defined. It can either be a laboratory measured transition wavelength, or better the wavelength of an observed stellar spectral line, to account for gravitational red shift and mean convective granulation blue shift. An ensemble of such lines is called a "Binary Mask", or BM.\\
The algorithm has other features:
\begin{itemize}
    \item it does not need a very accurate wavelength spectral calibration of all pixels/spectels. It requires though a great stability of the spectrometer between the two epochs.
    \item the pixels need not to be equally spaced and with a uniform size
\end{itemize}
However, the algorithm does not work properly when RV$_1$ and RV$_2$ are quite different, which occurs because of the Earth’s 30 kms$^{-1}$ orbital motion, inducing a shift on the detector of many pixels: the formula \ref{eq:deltaa} may not be used any more, because the same pixel samples at epoch E$_2$ a quite different portion of the spectrum, with a different slope, than at epoch E$_1$ (Figure \ref{fig:fig1}). The typical width of a spectral line is of the order of 3-10 kms$^{-1}$; see Figure \ref{fig:fig1}, with simulation of +2 and +15 kms$^{-1}$ shifts. However, this difficulty can be circumvented by using the formula (1), not applied to a pixel and trying to determine the shift in pixel units, but to a spectel, a wavelength element, and to use a wavelength scale in the frame of reference of the star target. An approximation of this requirement may just to correct the wavelength scale of the measured spectra from the BERV (Barycentric Earth Radial Velocity). This was done for instance for exemple in \citet{Dumusque2018}, who applied the Connes formula in a line-by-line template matching analysis. We note that none of the two formulas of Connes is giving an estimate of RV.

\section{The Cross-Correlation Function (CCF) algorithm to determine a radial velocity RV.}
\label{sec:ccf}

\citet{Baranne1996} designed the spectrometer ELODIE, partially inspired by the work of \cite{Connes1985}, and stimulated by Michel Mayor. ELODIE is a cross-dispersed spectrometer and CCD. Installed at Observatoire de Haute Provence, it allowed rapidly the discovery of 51 Pegasi b, the first exo-planet found around a Sun-like star, detected by the RV indirect method \citep{Mayor1995}. Earlier on,  planets had been detected around a pulsar \citep{Wolszczan92}.
\\
In \citet{Baranne1996} there is a description of the CCF algorithm applied to the series of stellar lines. The stellar spectrum intensity is digitized on pixels, each with an assigned wavelength. A box-car shaped "emission" line is 0 everywhere, and 1 in a small wavelength interval (for instance one pixel) centered on the absolute wavelength chosen for the transition responsible for the stellar line. 
The correlation function is computed on a fixed grid of potential absolute RV values (in kms$^{-1}$). Each point of the CCF is the portion of the stellar spectrum which is included in the box-car. Usually the CCF is fitted by a Gaussian, and the wavelength of the Gaussian minimum reveals a "proxy" of the dynamical absolute radial velocity  of the star.  A series of "emission" lines constitutes a Binary Mask. Piling-up together all individual CCFs of one optical Echelle order of the spectrometer, to get one single CCF per order has the great advantage to average out some artefacts: e.g., PRNU (Pixel-to-pixel Response  Non Uniformity); optical fringes caused by the optics of the spectrometer, imprinted on the observed spectrum; the blaze effect, which results in a bias when the CCF on one single line is fitted by a Gaussian.\\
In a word, the piled-up CCF algorithm is quite robust to several artefacts. Indeed, any pattern which is fixed in the wavelength frame of the spectrometer, while the star spectrum is moving by up to $\pm$ 30 kms$^{-1}$, will be detrimental (producing a bias on RV) for all methods which compare directly the motion of the spectrum, as the shift finding algorithm or others, like template matching . This is the case of several artefacts like PRNU, DCNU (Dark Current Non Uniformity), optical fringes, and micro-tellurics (large tellurics are just avoided by standard Binary Masks). For template matching,  when the template is built from many observations with full yearly excursion of BERV, it will somewhat smooth out artefacts.
Also, for CCF building some weights are usually assigned to each line of the Binary Mask \citep{Pepe2002}. For instance, \cite{Lafarga2020} use as a weight for each line the ratio between the contrast and the width of the line, where the contrast is the relative depth of the line and is given for each line of the official ESPRESSO BM. Then, all orders are combined together through the optimal combination of Gaussian errors, with a weight equal to 1/$\sigma_{ord}^2$. The uncertainty estimate for this order $\sigma_{ord}$ is given by the second formula of Connes applied to the CCF, as described in \cite{Boisse2010}. Another way to combine the CCFs from all Echelle orders is just to add them together to get a single CCF$_{tot}$ per exposure, since each of the order CCF has been obtained from the addition of all CCFs of the BM lines inside this order, already weighted for each line. This is indeed the case of the official ESO/ESPRESSO pipeline at VLT. While a good pipeline is supposed to correct many artefacts quoted above (PRNU, DCNU, optical fringes, micro-tellurics, stray light, etc.), one should still worry about the residuals after the correction of these effects by the pipeline. Indeed, \cite{Artigau22} have improved the RV data by eliminating outliers (resulting from residuals after pipeline corrections) with a Line-by-Line analysis of Barnard’ s star acquired with SPIRou spectrometer at CFHT.

\section{Proposed algorithm: applying the first formula of Connes to piled-up CCFs.}
\label{sec:proposed}

\subsection{Description of the new EPiCA algorithm.}
We propose to combine the optimal retrieval, photon noise limited, algorithm of Connes (equation \ref{equation8}) with the robustness of the piled-up CCF.  We call it the "CF1 to CCF" algorithm (CF1, Connes formula 1) or the EPiCA method. Instead of applying the 1$^{st}$ formula of Connes to a measured spectrum at two epochs, we apply it to the two CCF$_{tot}$ computed for those two epochs, yielding CFF1 and CCF2. The trick is that the CCF is not computed on the wavelength scale provided by the laboratory calibrated spectrometer, but on the wavelength scale produced after BERV correction (Barycentric Earth Radial Velocity). Indeed, in the solar system barycentric frame of reference, the radial velocity of a star is constant, except for the small variations induced by the potential presence of exo-planets. Today, conversions from geocentric to barycentric velocities have reached accuracy on the order of cms$^{-1}$ \citep{wright_eastman20} and are not a limiting factor. Therefore, there is only a small displacement of the two curves CCF1 and CCF2, suitable to use the Pierre Connes algorithm (equation \ref{eq:deltaa}) valid for small displacements. To the best of our knowledge, this method has never been used nor described before.\\
In practice, in the CCF fits files taken from the ESO archive the CCF matrix has 171 columns, while there are only 170 Echelle optical orders. The last column of the CCF matrix is an order-merged CCF$_{tot}$ produced by the pipeline itself. We have verified that this CCF$_{tot}$ is computed by adding together all CCF per order for the exposure. We applied the EPiCA algorithm to this order-merged CCF$_{tot}$ of the pipeline (last column numbered 170 in the code, where numbering starts at zero).  One particular exposure spectrum 1 is taken as a reference, providing CCF1$_{tot}$. Each other exposure 2 provides a CCF2$_{tot}$, and the 1$^{st}$ formula of Connes (Equ. 8) is applied, yielding the change of RV, DRV, between exposure 1 and exposure 2. The CCF2$_{tot}$ must be normalized to CCF1$_{tot}$ before applying Equ. 8 (both CCF1$_{tot}$ and CCF2$_{tot}$ must have the same integral). But note that all computations required for Equ. 8 use CCF2$_{tot}$ before its normalization to compute the weighting function W(i) and the quality factor Q (the “on the fly" quality factor and W$_{i}$ for each exposure). \\

\subsection{Validation: simulation of shift finding algorithm applied to an observed CCF.}
\begin{figure}
   \centering
     \includegraphics[width=0.5\textwidth]{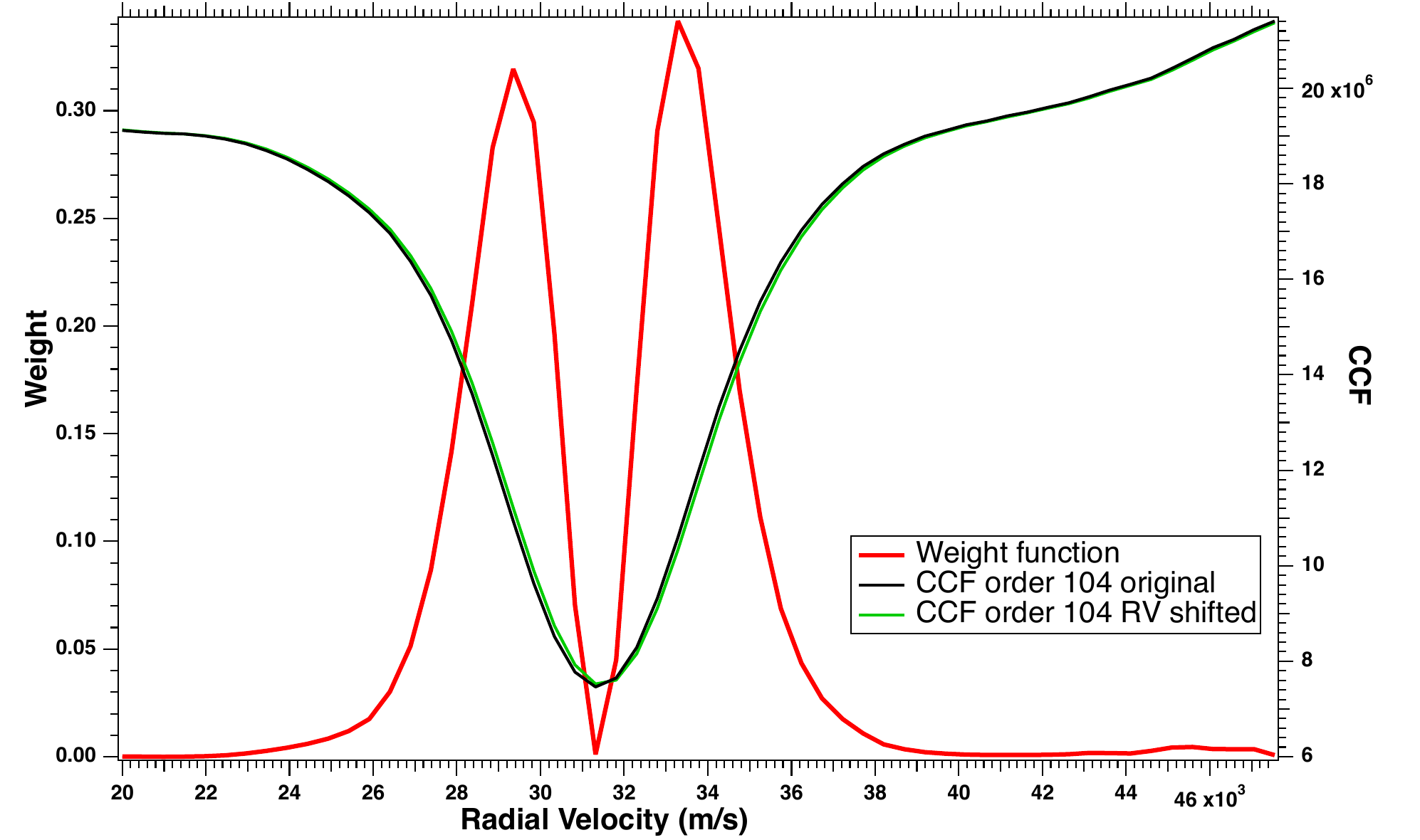}
  \caption{Illustration of the simulation used to validate the Connes shift finding algorithm and the error estimate. Black: CCF of original data (CCF1, right scale); green: CCF of simulated exposure shifted from original data (CCF2, left scale); red: weight function (left scale).}
              \label{fig:fig2}
    \end{figure}

Here we are testing the 1st and 2nd formulae of Connes (equations 8 and 9), applied to an observed CCF. For this simulation, we have used a series of 200 exposures taken on star HD 40307 \footnote{Prog.ID:0102.D-0346(A); PI: Bouchy} taken during the night from 24 to 25 December, 2018 (called night 24 elsewhere in this paper). We have added all 200 spectra together, spectel to spectel, for order 104, and computed the CCFs on a given grid V$_{rad}$ of radial velocities (figure \ref{fig:fig2}, black curve). The SNR for one spectel is about 1,000 for this reference spectrum S1. The wavelength scale WS1 was the spectrometer wavelength scale for exposure number 100. Order 104 (covering wavelength range 5523.8 – 5606.7 $\AA$) was selected because this order is not affected by tellurics, and the signal is rather high with one single exposure. The SNR for one point of the CCF is about 4,000.\\
Then, a synthetic spectrum S2 was derived from the stacked spectrum S1 just by modifying the wavelength scale, as if the star had changed its velocity by +100 ms$^{-1}$: same intensities but on a RV shifted scale WS2.   \\
In order to mimic what would be the values of spectrum S2 sampled on the grid WS1, we interpolate the spectrum S2 vs WS2 on all the points of the grid WS1.
The CCF2 of the spectrum S2 sampled on WS1 is done on the same grid V$_{rad}$ of radial velocities, yielding the green CCF curve of Figure \ref{fig:fig2}. 
Also displayed on Figure \ref{fig:fig2} (in red) is the weight attached to each point of the V$_{rad}$ grid (equation 8), according to the photon noise limit of Pierre Connes first formula, here applied to the CCF1.
The double-peak shape of the weight curve (in red) just shows the relative contribution (weight) of each point of the CCF to determine a shift: the weight is depending on the square of the derivative (w.r.t. the V$_{rad}$ grid):
\begin{equation}
    weight(i)=1/\sigma^2(i)=(\frac{\partial A_0(i)}{\partial sp})^2 \frac{1}{2A_n}
    \label{eq:weight}
\end{equation}
When the first formula of Connes is applied to the CCF1 and CCF2, a radial velocity change DRV of  99.613 ms$^{-1}$ 
is returned, while the second formula of Connes giving the uncertainty applied to CCF1 (or CCF2) is 0.366 ms$^{-1}$.
The retrieved DRV is near the + 100 ms$^{-1}$ simulated RV shift, with a difference from 100 ms$^{-1}$ near the nominal error bar. Therefore, this simulation exercise validates this new algorithm, and the uncertainty associated with it. It can be pointed out that, while applying a shift finding algorithm to the comparison of spectra as a function of wavelength includes an approximation (finding a shift instead of a stretch), applying the same shift finding algorithm to a CCF does not imply the same approximation, because the scale of the CCF is a radial velocity, and computing the CCF accounts exactly for the Doppler stretching.

\subsection{Basic differences between the two RV retrieval algorithms.}

The ESO/ESPRESSO pipeline algorithm for RV retrieval is to fit the curve CCF$_{tot}$ (similar to green and black curves on Figure \ref{fig:fig2}) by a symmetric Gaussian and assign as absolute RV value the position of the center of the Gaussian. Therefore, only one single information is derived from the full CCF$_{tot}$ curve. Actually, as shown by \cite{Gonzalez2020} for the Sun (integrated disc through observation of the Moon), the exact position of this symmetry axis of a Gaussian fit to a single solar line depends somewhat on the wavelength extent of the fitting domain, which shows that the spectral lines are not rigorously symmetric. Inasmuch as the curve CCF$_{tot}$ is an average image of all piled-up solar lines of the Binary Mask, it is likely that such a dependence also exists for the CCF$_{tot}$. Then, RV changes of a star are just computed by subtracting a constant from all RV measurements.\\
On the other hand, the Connes algorithm does not claim to determine an absolute value of RV for each spectrum and corresponding CCF$_{tot}$, but only the change DRV of RV between two spectra and their corresponding CCF$_{tot}$. Actually, each point of the CCF$_{tot}$ gives an estimate of DRV, which can be combined together in an optimal way (mathematically) by applying the weight function of equation \ref{eq:weight} (red curve of figure \ref{fig:fig2}). This is what we have done in the following. However, there is potentially the possibility to study separately various parts of the spectral lines (or their image trough the CCF$_{tot}$). In particular, the red side and the blue side could give different estimates of the RV change, if stellar activity deforms the spectral lines and CCF$_{tot}$ in a non-symmetrical way. 
In summary, it is clear that the ESO pipeline for ESPRESSO must give, for each observed spectrum, one single number, an estimate of the absolute radial velocity of the star (in the barycentric system), revealed by its Doppler effect ; and this is indeed what is provided by the present version of the ESO ESPRESSO pipeline when fitting CCF$_{tot}$ by a symmetric Gaussian. \\
When a change of RV between two spectra (taken at two epochs) must be evaluated, the simplest method consists of comparing RV$_1$ and RV$_2$ from the pipeline, and this is exactly what is done routinely during the search for exoplanets by most of scientific teams. 
However, the two spectra contain an enormous amount of information: they could be compared, in the extreme limit, spectel by spectel, with the Pierre Connes approach, each spectel giving an estimate of the DRV= RV$_2$-RV$_1$. This could be a way to detect some spectral regions which are affected by stellar processes changing the shape of the spectral lines (including variations of convective Granulation Blue Shift, GBS). Unfortunately, there are some artefacts spoiling this extreme approach: PRNU, DCNU not perfectly known and accounted for; optical fringes, which may change with the orientation of the telescope; uncorrected micro tellurics, to name a few. These outliers are well detected by the \citet{Artigau22} technique which explains its success in increasing the accuracy of RV. On the other hand, using CCF order-by-order and even CCF$_{tot}$, mitigates these artefacts by averaging them out: the CCF method is quite robust. 
With the proposed EPiCA method (applying the Connes approach to the CCF$_{tot}$), there is a hope to combine the robustness of the CCF approach (which combines many lines together, and partially smoothes out some artefacts) and the sophistication of the Pierre Connes' approach.

We note two other interesting studies that are using also CCF$_{tot}$, with the aim (the same aim as us) to try to discriminate planetary reflex Doppler shifts from stellar variability. \cite{Simola19} are adding one more parameter in the Gaussian fit to the CCF, the skewness (asymmetry). \cite{Collier21} use the autocorrelation function (ACF) and departures from an average of all ACF, which requires a lot of observations. Here we explore an alternate route, by studying in detail how the CCF$_{tot}$ behaves, with the help of the Pierre Connes shift finding algorithm and the full details provided by the DRV curves.

\section{Application to HD 40307 ESPRESSO data and 3 planets model.}
\label{sec:application}

\subsection{Overview.}
Ideally, in order to compare two methods yielding changes of RV, one should use a planet-less star with a constant RV, and compute the standard deviation of a series of measurements of RV changes around the mean which should be 0. On the other hand, we wished to use ESPRESSO data which provide an absolute wavelength scale of excellent quality \citep{Pepe2021}, possibly the best in the world, and extensive series of measurements. Looking at the available ESPRESSO data, we noticed a quite exceptional series of 1151 exposures spread over 7 days of star HD 40307. HD 40307 is a K2.5V type star with visual magnitude 7.147 and distance 13 parsec. It was observed during 6 nights from 22 December to 28 December 2018 by ESPRESSO. The series of exposures were made in HR mode which resolution varies between 100 000 and 160 000 across the spectrum. It had originally be planned to search for an asteroseismology signal (finally, not found yet on this star), but has the advantage of providing a unique way to get an estimate of the true error made on RV changes, based directly from data on the dispersion of RV change residuals to a linear best fit during a short period of time. For this asteroseismological campaign\footnote{Prog.ID:0102.D-0346(A); PI: Bouchy}, the exposure time was set to 30 seconds for all 1151 exposures, and the sampling period was about 78 seconds.\\
The star HD 40307 harbors a multi-planetary system. It was reported by \cite{Maoyr2009} that it  hosts 3 super-Earth-type planets, with orbital periods P$_b$=4.311 days, P$_c$=9.6 days and P$_d$=20.5 days. Later, \cite{Tuomi2013} independently analyzed HARPS publicly available data and concluded to the presence of 3 additional exoplanets with periods: P$_e$=34.62 days, P$_f$=51.76 and P$_g$=197.8. \cite{Diaz2016} confirmed the 3 planets found by \cite{Maoyr2009} and planet f from \cite{Tuomi2013}, but cast doubt on the presence of planets e and g. The most recent analysis made by \cite{Coffinet2019} confirm the findings of \cite{Diaz2016}: they found 3 exoplanets from \cite{Maoyr2009} and also the additional planet f from \cite{Tuomi2013}, but as well as \cite{Diaz2016} cast doubt on the presence of planets e and g. After subtraction of a long-term signal, which probably represents a magnetic cycle of the star, the signal around the period of 200 days has a very low significance and the signal around 34.6 days is absent.\\ 
In our analysis, we considered only the 3 planets b, c, d, and excluded planet f, because its period of 51.76 days is considerably longer than the short duration (7 days) of the time series and would not make any significant difference in a model fit to data.  
\subsection{Comparison of the two RV time series obtained with the two methods.}
\begin{figure*}
 \centering
  \includegraphics[width=0.97\textwidth]{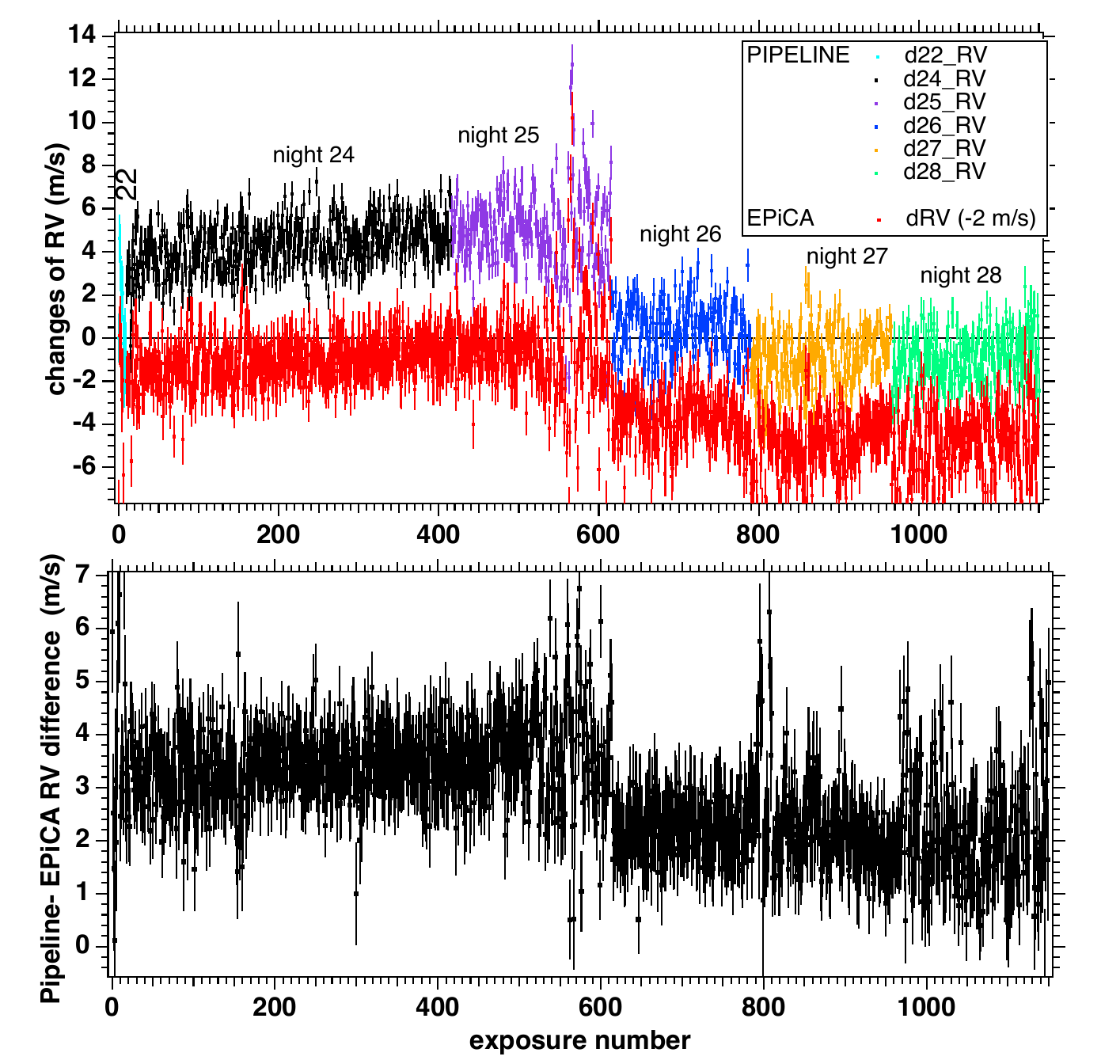}
  \caption{Top: Comparison between the RVs derived from the Gaussian fit and the EPiCA method: the data (RV changes in ms$^{-1}$) are plotted as a function of exposure number. The pipeline RV are plotted with their error bar, with one different color for each night. They were derived by subtracting from all pipeline RV data contained in each spectrum fits file the value of the first exposure: 31.3668 kms$^{-1}$. Red points are obtained by the EPiCA method applied to the CCFs of the current exposure and the first exposure of night 24 (exposure n$^o$10 when all nights are concatenated), taken as a reference. A constant 2 ms$^{-1}$ was subtracted for clarity. Bottom: Differences between the two time series of RV changes: Gaussian fit - EPiCA (CF1 applied to CCF$_{tot}$). This difference shows an abrupt decrease ($\sim$1.3 ms$^{-1}$) after exposure n$^o$615, corresponds to transition from night 25 to night 26, which remains after that.}
    \label{fig:fig3_4}
\end{figure*}

We now compare the two RV time-series: the "official" ESPRESSO, containing the RV assigned to each exposure recording one spectrum of HD 40307 (the pipeline series), and the time series that we obtained from our new EPiCA method, CF1 applied to CCF$_{tot}$. They are plotted on Figure \ref{fig:fig3_4}, not as a function of time, but as a function of exposure number; the data are concatenated. From the pipeline RV data was subtracted a constant radial velocity of 31.3668 kms$^{-1}$, while the EPiCA series displays the RV change between each exposure and exposure n$^o$10 taken as an arbitrary reference. Therefore, the overall offset between the two time series is arbitrary and has no particular significance.\\
Both time series show a rather sharp decrease after exposure 615, which is the last one of the night 25 (in short for 25 december). This is due to the presence of planets, as we shall see later. 
The offset between the two series is partially due to the difference of reference exposure: n$^o$0 for ESPRESSO pipeline, n$^o$ 10 (first of night 24) for Connes CF1 applied to CCF$_{tot}$. But the offset is not constant, as can be seen when plotting the difference between the two time series (Fig. \ref{fig:fig3_4}). There is a change of $\simeq$1.3 ms$^{-1}$ between nights 25 and 26.

Therefore, one of the two methods is giving some wrong results by about 1.3 ms$^{-1}$ (or may be both of them), which do not correspond to a true change of the dynamical radial velocity dR/dt. We suspect that this is the effect of stellar processes, and more precisely of granulation blue shift (GBS) changes, with a time scale of a few hours, together with the fact that both methods are differently sensitive to stellar processes. 

\subsection{ Comparison of RV dispersion over one night from a linear fit.}
One criterion for RV retrieval method comparison is the use of a time series of measurements when the star is not moving at all. The precision of the RV method is quantified by the mean error, equal to the standard deviation $\sigma$ (the square root of the variance) of the series of measurements (this is the definition of the precision). In our case of the HD 40307 time series, the star is moving because of planets, and RV is changing. However, we may restrict the time series to a particular night and a limited duration in such a way that the RV variation is small and close to linear. In this configuration, we may accommodate with planet-induced star motion by subtracting a linear fit and computing the standard deviation of the residuals, as a quality indicator of the method. Actually, this linear fit may also include spurious RV variations due to stellar variability affecting the Doppler shift of the disc-integrated spectrum. \\
The first night contains only 10 measurements and is not suitable for this statistical exercise. The third night (25 December) displays a significant number of outliers, affecting the second half of the 200 measurements (figure \ref{fig:fig3_4}). This behavior has been assigned to an instability of the cooling system of the ESPRESSO blue detector \citep{Figueira2021}. Therefore, the night of 24 December containing 406 measurements was selected for further comparison of the two methods. \\
As said above, we compare on Figure \ref{fig:fig3_4} the EPiCA results with the results from (symmetric) Gaussian fits to CCF$_{tot}$  (the RVs from the pipeline). Although not easily visible by eye, there is a gain (i.e. a reduction) in RV dispersion $\sigma$ in the case of the EPiCA method. 
The cloud of data points can be seen on figures \ref{fig:fig5} and \ref{fig:fig6} at night 24 and day 2.2. 

\begin{figure}
    \centering
    \includegraphics[width=0.5\textwidth, height=0.4\textwidth]{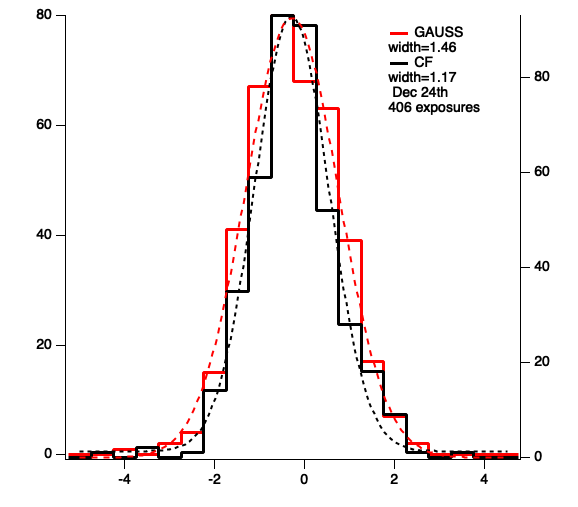}
    \caption{Histograms of RVs obtained from Gaussian fit (red) and EPiCA (black) for the 406 exposures of night Dec24, after removal of linear fits to each series. Gaussian widths of histograms are indicated for each method. The EPiCA histogram is narrower than the Gaussian fit histogram.}
    \label{fig:fig12}
\end{figure}

In order to compare quantitatively the two methods we made a linear fit to the 406 values of RVs of night December 24th for both measurement methods and removed a linear trend (due to planets and/or stellar processes) to keep only the residuals, and compute the dispersion around the average value. We made two histograms of the residual RV values with the same bins and compared them, as shown on Figure \ref{fig:fig12}. The figure clearly shows the gain in precision with the EPiCA method, quantified by the widths of Gaussian fits to the two histograms. For this series of exposures, the semi-width of the Gaussian (at 1/e) is reduced with the EPiCA method w.r.t. the pipeline, from 1.46 to 1.17 ms$^{-1}$, corresponding to a reduction of the dispersion (the actual mean error on RV changes measurements) from $\sigma$=1.03 to 0.83 ms$^{-1}$, a significant 20\% improvement of the precision.

The result is not very surprising, since P. Connes demonstrated that his first formula is the optimal way to measure Doppler shifts, by taking into account the totality of information contained in the two spectra (or here, the CCF$_{tot}$). On the other hand, as we will discuss below, the hypothesis underlying the Connes formulation is an absence of variability of the shape of the spectra (or here CCF$_{tot}$), except for a single shift. Only pure Doppler shifts  are allowed. We interpret this decrease of the RV dispersion during a short duration as an evidence for the absence of strong variations of the stellar line shapes, at least of the absence of shape variations strong enough to cancel the benefit of the application of the CF1 formulation.   


\subsection{A 3-planets best fit to the official pipeline ESPRESSO RV time series.}\label{bestfit}

\begin{table*}[h!]
\begin{tabular}{|l|l|l|l|} 
  \hline
&HD 40307b&HD 40307c&HD 40307d\\
\hline
   Amplitude and error [ms$^{-1}$]& 1.84$\pm$ 0.14 &  2.29$\pm$ 0.13 & 2.31$\pm$ 0.14\\ 
  \hline
Nominal period and error [day]&4.3114$\pm$0.0002&9.6210$\pm$0.0008&20.412$\pm$0.004\\
\hline
 Zero crossing time T$_0$ at epoch 2008 (bjd-2.4e6)&54562.77$\pm$0.08&54551.53$\pm$0.15 &54532.42$\pm$0.2\\ 
  \hline
m$_{2}$sini (M$_{E}$) (Mayor et al., 2009) &4.2 & 6.9 & 9.2\\
 \hline
 \end{tabular}
\caption{Characteristics of three planets in HD 40307 system from HARPS 2008 campaign \citep{Coffinet2019}}
\label{tab:HD40307planets}
\end{table*}

We now examine a second quality indicator of an RV changes retrieval method, totally independent of the first one, namely the true error computed from dispersion,  discussed in the previous section. Our goal is to investigate whether or not the EPiCA method may be less sensitive than the Gaussian fit to spurious changes of RV due to stellar activity. The idea is that the RV changes due to planets do obey strictly to Kepler's laws, while RV changes due to stellar activity are more random (but still stochastic, \citet{Meunier2021, Meunier23}). Therefore, we may produce a quality indicator indicating how well a series of measurements reproduce a pure Kepler's law model of RV changes. To do so, we have used the entire HD 40307 time series. For each method, we have determined some best fit parameters of the 3 planets system, and computed the absolute difference between data and best-fit model as the second quality indicator. In Table \ref{tab:HD40307planets} we summarize some parameters of the 3 planets detected around HD 40307.

\subsubsection{Best-fit strategy and result of RV official pipeline ESPRESSO RV time series.}\label{best-fit}

\begin{figure}
\centering
    \includegraphics[width=0.5\textwidth]{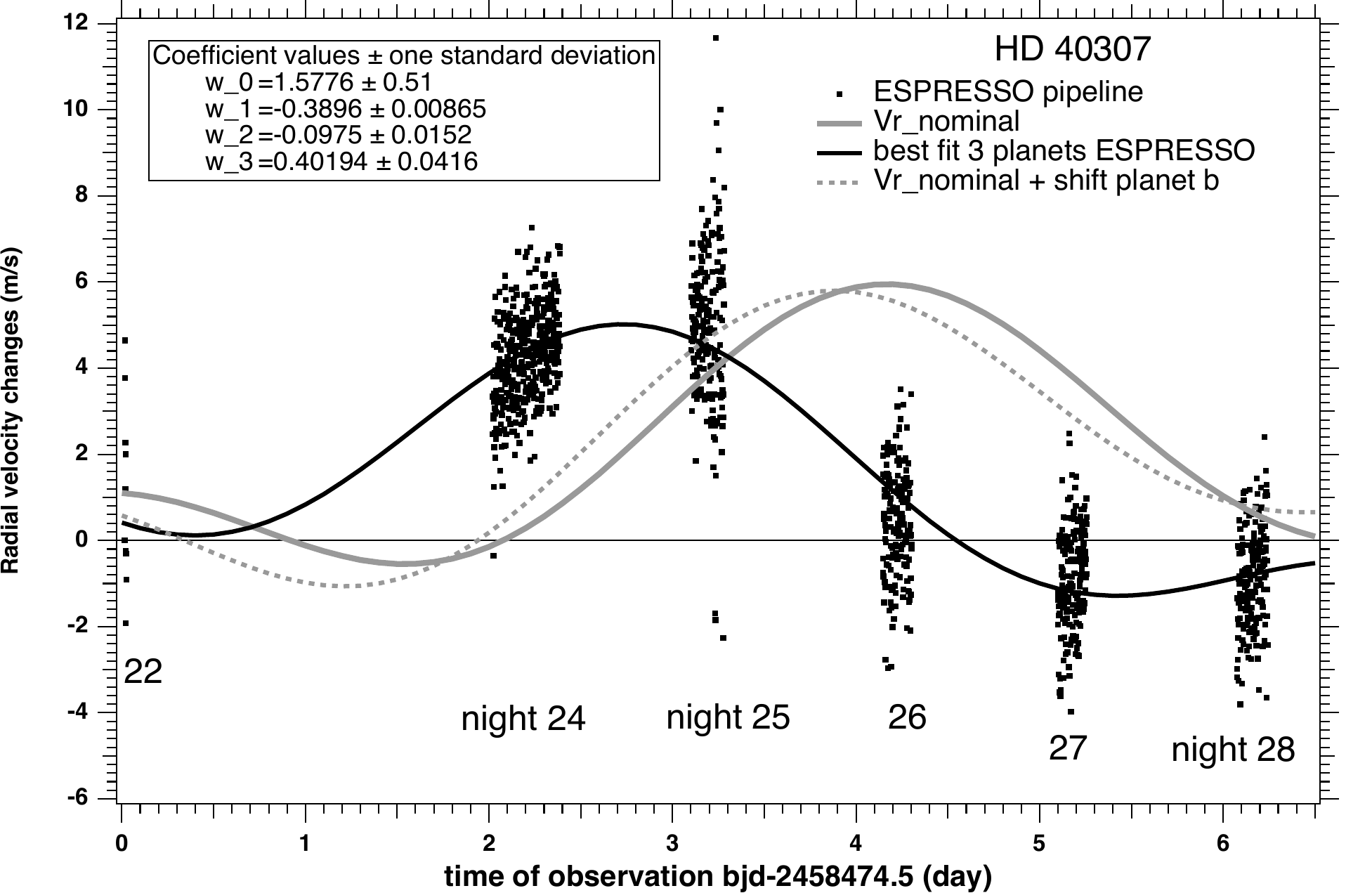}
    \caption{Three-planets modeling: the black dots represent the time series of changes of RV values contained in the official pipeline, obtained by subtracting the RV value of the first exposure of the first night. The thick pale grey solid line is the extrapolation of the sum of the 3 sine waves as determined form 2008 HARPS data. It does not fit well the data, because the extrapolation over 9 years is sensitive to small errors in the periods. The dashed grey lines is the same as the solid grey line, but with  planet b period shorter by 0.0004 d, corresponding to twice the claimed error bar for the period of planet b. The solid black line is the best fit to the pipeline RV data by adjusting the phases of the 3 planets.}
    \label{fig:fig5}
\end{figure}

Our first analysis is done only on ESPRESSO 2018 data, without archive HARPS data. ESPRESSO data set consists only of 6 days, so it is not possible to detect a signal from the 4$^{th}$ (HD 40307 f), since it would induce only a too small drift over 6 days. We have assumed that planets are on circular orbits as it is mentioned in Table 9 from \citet{Diaz2016}.\\
Therefore, we kept fixed the amplitudes and periods of 3 planets at their most recently published values from \citet{Coffinet2019}, where the authors corrected the HARPS wavelength calibration from the CCD stitching, a gap in blocks of pixels composing the full CCD. We know the time T$_0$, the origin of the sinusoidal wave at epoch 2008 from the \href{http://exoplanet.eu }{exoplanet.eu} (see Table \ref{tab:HD40307planets}). With these time values of ascending zero crossing in March-April 2008, we can extrapolate in time all 3 sine waves with the defined periods up to the epoch of 2018 observations, getting the curve RV$_{nominal}$ (pale grey, thick solid line) on Figure \ref{fig:fig5}, compared to ESPRESSO data. We see a major discrepancy, easily explained: an  error on any of the periods, propagated over 10 years, will make RV$_{nominal}$ not representing well the data of 2018. Also represented with a dashed grey line is the extrapolation with a period of planet b shorter by 0.0004 d, twice the claimed error bar. We are still far away form a reasonable fit.

\begin{figure}
\centering
    \includegraphics[width=0.5\textwidth]{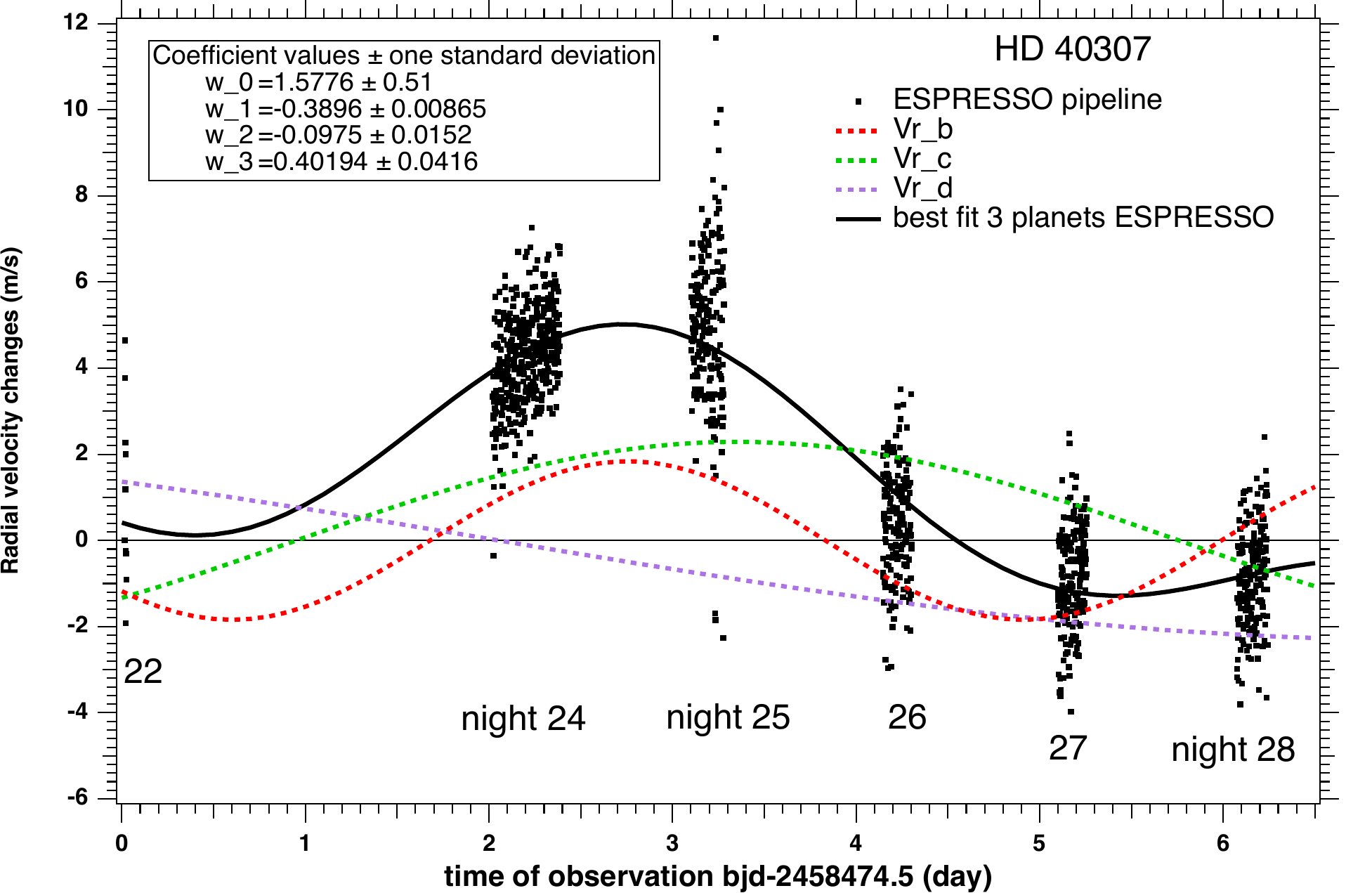}
    \caption{Same as figure \ref{fig:fig5} with individual contributions of the planets. The 3 planets best fit curve is in black (solid line), while the dashed lines are each of the 3 sine waves representing the 3 planets : b (red), c(green), d (violet). The best fit curve is the sum of the three planets plus a constant w$_0$= 1.5576 $\pm$ 0.515 ms$^{-1}$.} 
    \label{fig:fig6}
\end{figure}

Therefore, we have to adjust the exact times of ascending zero crossing near the time of observations,  for each of the sine functions in order to get a good fit of data.
Actually, we take as unknown in the fitting process the exact phase of each of the sine function at a reference time (BJD$_{ref}$= 2458474.5), taken as the time 0 of the plot of Figures \ref{fig:fig5} to \ref{fig:fig11}. It is adjusted by a classical Levenberg-Marquardt scheme within IGOR software. 
We also have to let free a constant offset parameter w$_0$, added to the three sine functions, since we deal only with variations of RV. 

On Figure \ref{fig:fig6} is plotted the best fit curve, in black, while the 3 sine waves for the 3 planets are represented by dashed lines around the zero line. It is clear that the drop in RV between night 25 (day 3.2) and the next night 26 is mainly due to planet b, with smaller contributions from planets c and d. 

\subsubsection{Refining the periods by comparing 2008 and 2018 data}\label{best-fit}
For each of the sine curve, knowing the phase at BJD$_{ref}$ allows then to compute the time T$_{2b}$ (respectively T$_{2c}$, T$_{2d}$) of start of the period ($\sin$=0 and becoming positive) just before BJD$_{ref}$. The values are displayed in Table \ref{tab:HD40307planetb} with their uncertainties. \\
We find in the literature that a similar sinus=0 crossing happened for planet b at T$_{0b}$ = 2454562.77$\pm$ 0.08 BJD, on April 5, 2008. Therefore, we may determine a precise value of the average orbital period of planet b between epochs April 2008 and December 2018, knowing that it must be an integer number of periods between T$_{0b}$ and T$_{2b}$. The elapsed time difference T$_{2b}$-T$_{0b}$= 3909.0964 $\pm$ 0.088 days. Here we take as the uncertainty the quadratic sum of the two uncertainties, respectively 0.08 and 0.037 days for T$_{0b}$ and T$_{2b}$.
Dividing the elapsed time difference by the nominal period 4.3114 gives a number of orbits of planet b K$_{per}$=906.688 (Table \ref{tab:HD40307planetb}), while it should be an integer number. We designate by K$_1$ and K$_2$=K$_1$+1 the two integer numbers encompassing K$_{per}$.\\
With 907 full orbits between T$_{0b}$ and T$_{2b}$, the period is found to be P$_{b907}$ = 4.30992 $\pm$ 1.0 10$^{-4}$ days, while with 906 full orbits, the period is P$_{b906}$ = 4.31467 $\pm$ 1.0 10$^{-4}$  days (Table \ref{tab:HD40307planetb}). These two values encompass the original value of the period found at discovery announcement 4.3115 $\pm$ 0.0006 and the more recent value Per$_b$=4.3114$\pm$ 0.0002. Therefore we favour the P$_{b907}$ which is about 2.2 times nearer this value Per$_b$=4.3114 than P$_{b906}$. Still, this new accurate value P$_{b907}$ is different from the nominal 2008 period Per$_b$=4.3114$\pm$ 0.0002 by 0.0015 d, outside the claimed error bar err$_{Per_b}$=0.0002 by a factor of 7. The planet b may have changed its average period from interactions with other planets between 2008 and 2018. Alternately, the uncertainty of 0.0002 d claimed for 2008 data was underestimated, but some arguments developed in Appendix B suggests that this is unlikely the case. Actually, similar results are found with both methods applied to 2018 observations, Gaussian fit (pipeline) or EPiCA (see next section). This is a potentially very interesting result, because it could reveal the influence of other planets on the orbit of planet b. However, since it is irrelevant to the comparison between the two RV methods, we defer this subject to Appendix B for further discussion. 

The same procedure was done for planets c and d, and the results are in Tables \ref{tab:HD40307planetc} and \ref{tab:HD40307planetd} respectively. Clearly, K$_{per}$ is nearer an integer number than for planet b, and the most likely values are K$_{per}$=407 for planet c, and K$_{per}$=193 for planet d.\\

\subsection{A 3-planets best fit to the EPiCA derived time series.}\label{threeplanets}
\begin{figure}
    \includegraphics[width=0.5\textwidth]{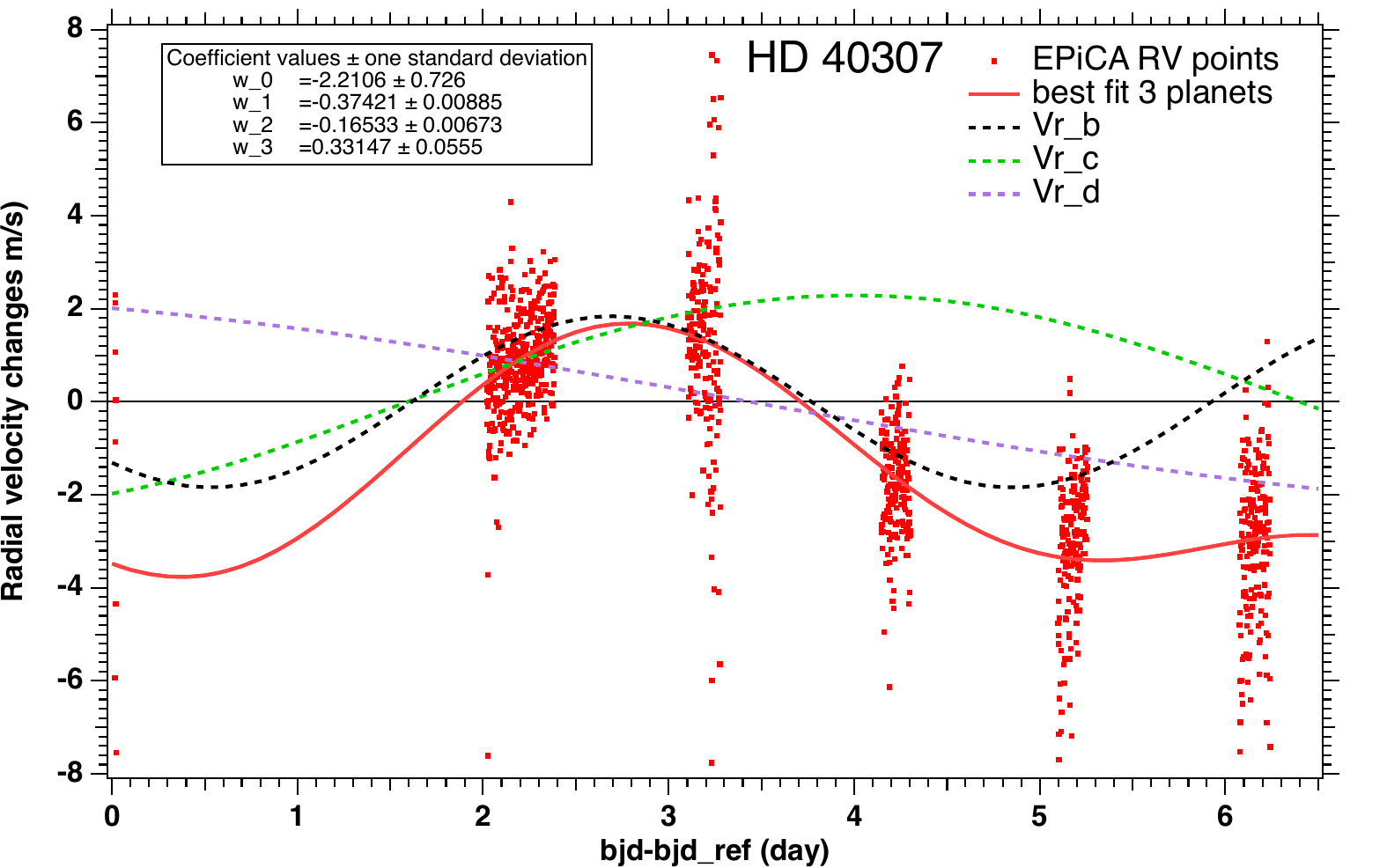}
    \caption{Same as figure \ref{fig:fig6}, but for the time series obtained with EPiCA (red dots), taking the first exposure as the reference. The best fit curve is in red (solid line), while the dashed lines are each of the 3 sine waves representing the 3 planets : b (black), c(green), d (violet). The best fit curve is the sum of the three planets plus a constant w$_0$= -2.2106 $\pm$ 0.726 ms$^{-1}$.}
    \label{fig:fig7}
\end{figure}

We want to find out if the use of EPiCA method (CF1 applied on CCF$_{tot}$) is able to produce an improvement in the case of a planetary system. To do so, we applied exactly the same fitting methods as in section \ref{best-fit}, this time by replacing pipeline results by EPiCA results. Figure \ref{fig:fig7} illustrates the fit by a model with 3 planets to data computed by EPiCA method (CF1 on CCF$_{tot}$).

\begin{figure}
    \includegraphics[width=0.5\textwidth]{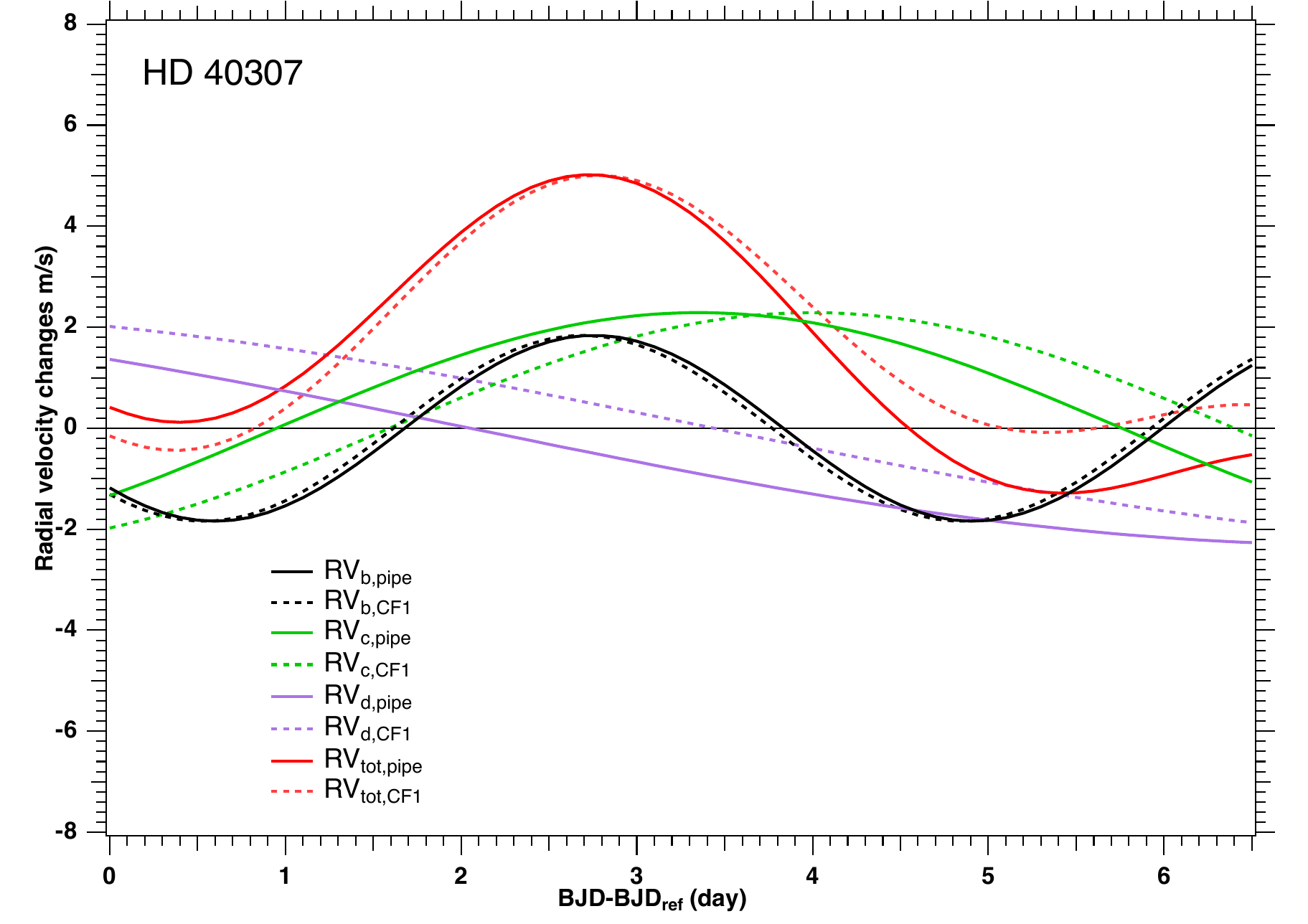}
    \caption{Comparison of best fit models to the two time series. Solid lines are for the ESPRESSO pipeline.  Dashed lines are for the EPiCA method. The red curve is for the total of 3 planets (+ the offset w$_0$), while planets b, c, d have different colors. }
    \label{fig:fig8}
\end{figure}

On figure \ref{fig:fig8} are compared the best fit curves for the 3 planets obtained with the two methods, pipeline and EPiCA . While there is not much difference for planet b, there is some substantial time shift for planets c and d. 

The same exercise to compute the period from the elapsed time between two zero crossings (from negative to positive) of the sine wave was done for the best fit to RV data retrieved with EPiCA method. The results (also shown in Table \ref{tab:HD40307planetb}, \ref{tab:HD40307planetc}, \ref{tab:HD40307planetd} for the two methods are not very different. For planet b, and with the EPiCA method we find the same period offset as the nominal period found at 2008 epoch. We have summarized in Table \ref{tab:newperiods} the new results of the periods of the 3 planets both for the best fit to pipeline data and EPiCA data. The results of both methods are similar. We have also indicated the difference of period between the nominal (old) value, and the new value found from the CF1 method. As mentioned above, this difference for planet b is 7.5 times larger than the previously claimed error bar. For planets c and d, the difference is 2.5 and 2 larger than the claimed error bar. In summary, joining the 2008 and 2018 data is providing more accurate periods than those which were determined with 2008 observations only, and this is valid for both methods. 
\begin{table*}[h!]
\begin{tabular}{ |c|c|c|c|} 
  \hline
&HD 40307b&HD 40307c&HD 40307d\\
\hline
   Amplitude and error [ms$^{-1}$]& 1.84$\pm$ 0.14 &  2.29$\pm$ 0.13 & 2.31$\pm$ 0.14\\ 
  \hline
Nominal period and error [day]&4.3114$\pm$0.0002&9.6210$\pm$0.0008&20.412$\pm$0.004\\
\hline
 New period and error [day] from pipeline data &4.3099 $\pm$0.0001&9.6174 $\pm$0.0005&20.383$\pm$0.0045 \\ 
  \hline
   New period and error [day] from CF1 method &4.30984$\pm$0.0001&9.619 $\pm$0.0004&20.4194$\pm$0.0006\\ 
     \hline
   Difference nominal- CF1 [day]  &0.0015&0.002&0.007\\ 
   \hline
\end{tabular}
    \caption[Characteristics of three planets in HD 40307 system from HARPS 2009 campaign. New periods are computed based on pipeline data and on CF1 method results]{Characteristics of three planets in HD 40307 system from HARPS 2009 campaign. New periods are computed based on pipeline data and on CF1 method results.}
\label{tab:newperiods}
\end{table*}

\subsection{Comparison of the residuals to 3-planets best fit between the two methods.}
\begin{figure}
    \includegraphics[width=0.5\textwidth]{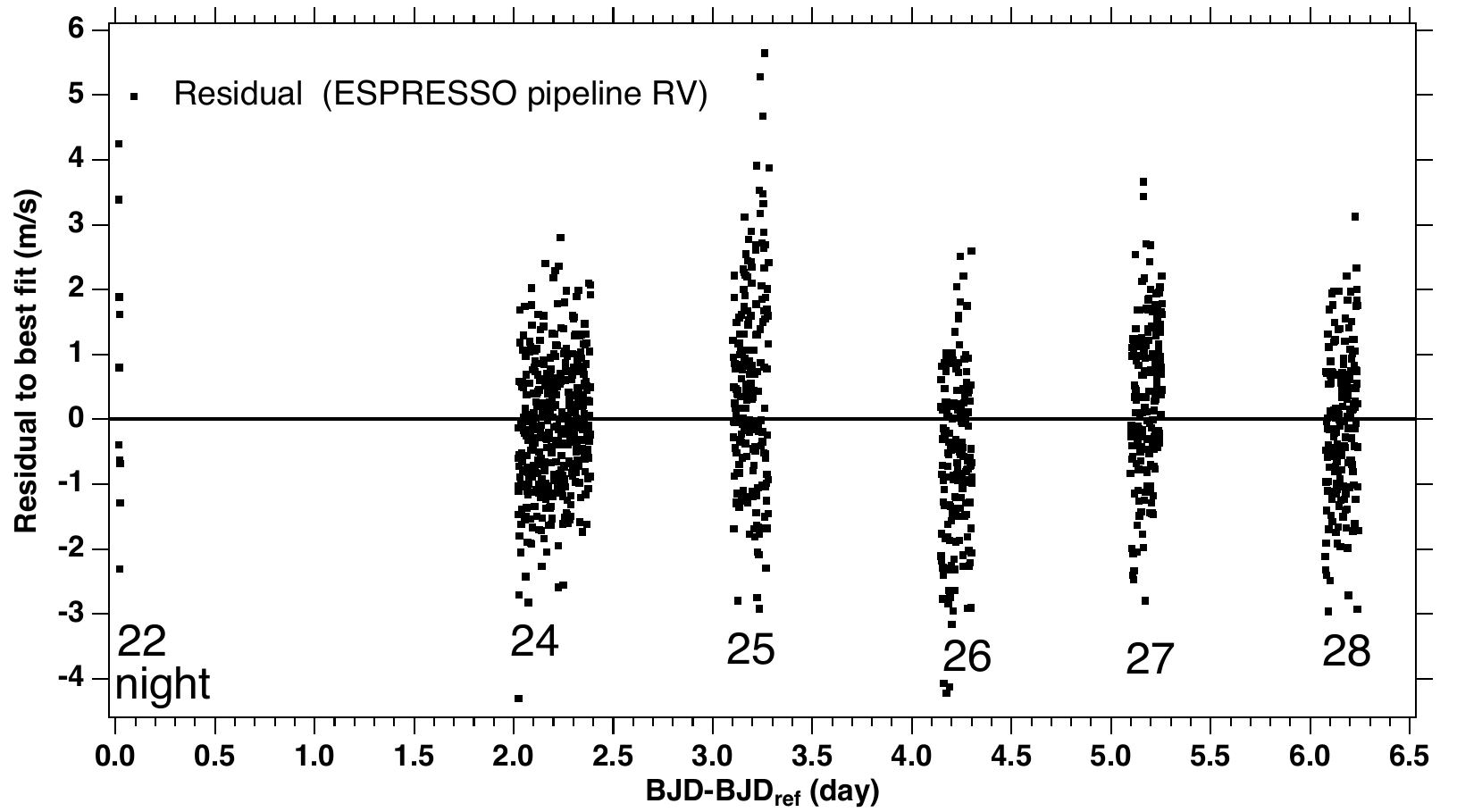}
    \caption{Residuals after subtraction of the best fit model from the $\delta_{RV}$ official Gaussian fit (black points). It seems by eye that the average of groups of points do not always lie on the zero line. There is a bias that depends on the day of observation (numbers represent the night of December 2018 when was observed HD 40307.) }
    \label{fig:fig9}
\end{figure}

\begin{figure}
    \includegraphics[width=0.5\textwidth]{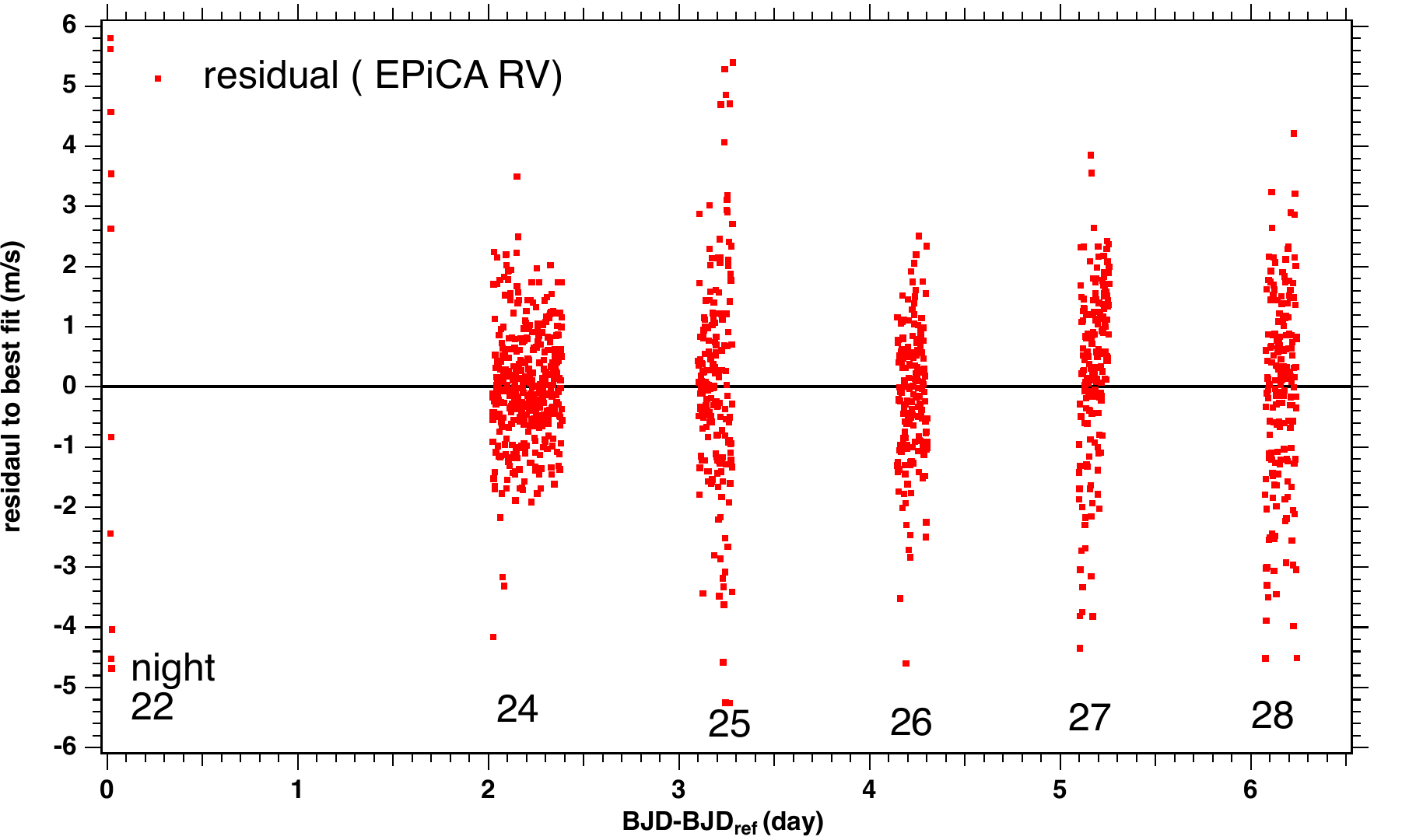}
    \caption{Residuals after subtraction of the best fit model from the CF1 to CCF$_{tot}$ time series (red points). It seems by eye better aligned on the Zero line (solid black line) than for the $\delta_{RV}$ time series of figure \ref{fig:fig9}.}
    \label{fig:fig10}
\end{figure}

On figure \ref{fig:fig9} are represented (black points) the residuals ($\delta_{RV}$ time series after subtraction of the best fit model) for the RVs from the pipeline (Gaussian fit), and similarly (red points) on figure \ref{fig:fig10} the residuals for the EPiCA method. By eye, it seems that the EPiCA method is better aligned on the Zero line (solid black line) than for the $\delta_{RV}$ time series from the pipeline. 

\begin{figure}
    \includegraphics[width=0.5\textwidth]{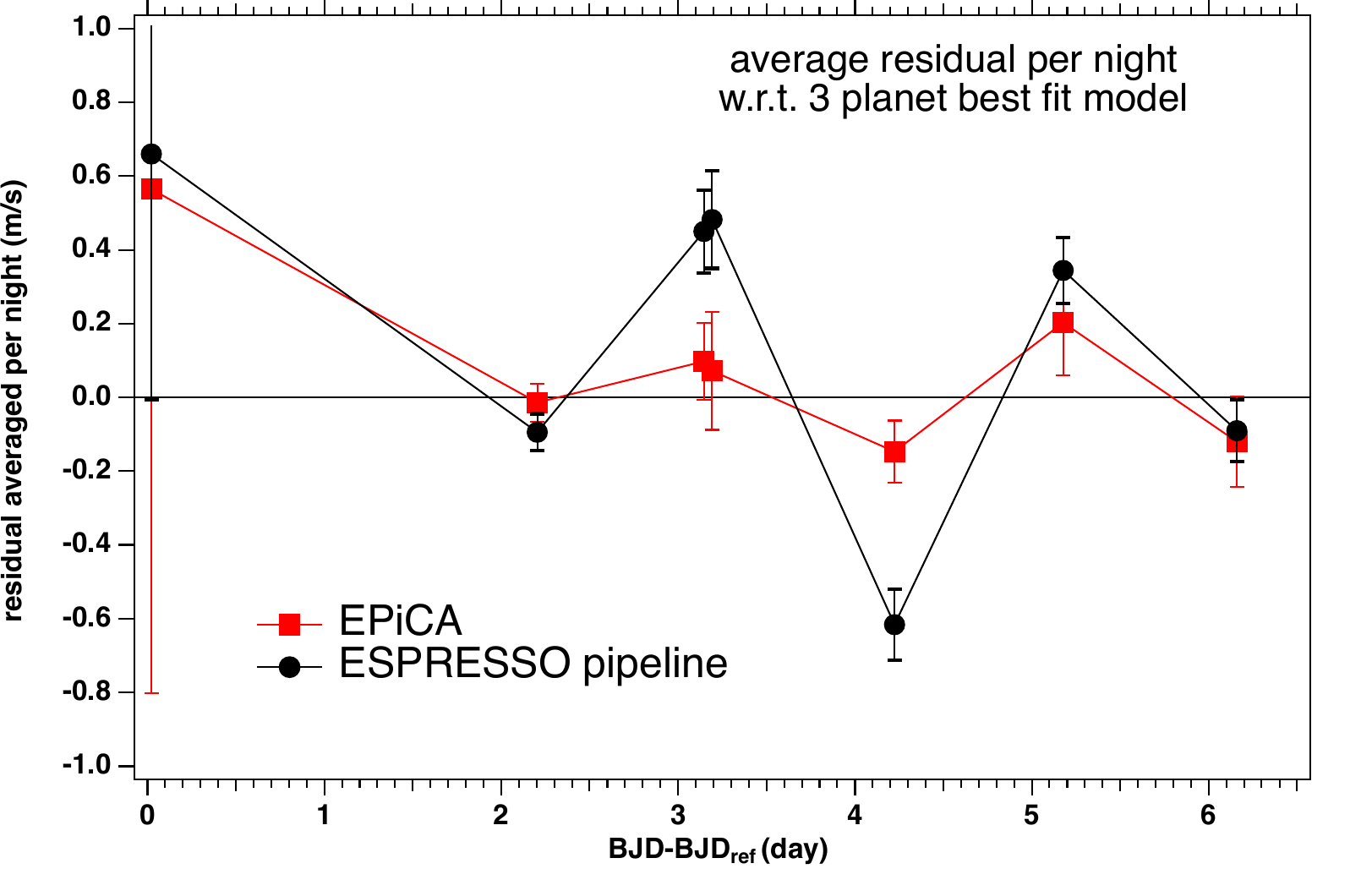}
    \caption{Mean residual for each night as a function of time (BJD-BJD$_{ref}$). Black: Gaussian fit. Red: EPiCA method (CF1 on CCF$_{tot}$). The error bars are computed as the standard deviation divided by sqrt(number of exposures). The two points for night 25 (around 3.2 days) are not independent.}
    \label{fig:fig11}
\end{figure}

This visual impression is confirmed by the following statistical analysis, summarized in Table \ref{tab:residuals} and on figure \ref{fig:fig11}.
The first night (December 22) contained only 10 exposures, and is not in Table \ref{tab:residuals}. 
The second night (24) is the longest one and contained 406 exposures. It shows the smallest standard deviation, quite similar for the two methods. The third night (200 exposures) had a specific problem. An instrumental noise due to a blue detector instability spoiled somewhat the second part of the night (visible on all figures with data, residuals, and standard deviation). This is why we computed the standard deviation and mean residual per night for both the first 100 exposures only and for the 200 exposures of the night, as two separate entries for night 25 in Table \ref{tab:residuals}.  

\begin{table*}[h!]
\begin{tabular}{|c|c|c|c|c|c|}
  \hline
Date&N$^o$ of exposures, exposure N$^o$&S$_{dev}$ pipeline&S$_{dev}$ CF1&Residual pipeline&Residual CF1\\
\hline
24& 406	[10,415]&	1.014&	1.03&	-0.094&	-0.014\\
\hline
25&100	[416,515] (*)&	1.13&	1.04&	0.45&	0.098\\
\hline
25& 200	[416,615] (*)&	1.87&	2.26&	0.482&	0.072\\
\hline
26& 174	[616,789]&	1.26&	1.11&	-0.616&	-0.147\\
\hline
27&176	[790,965]	&1.19&	1.91&	0.344&	0.203\\
\hline
28& 185	[966,1150]	&1.13& 	1.68&	-0.09&	-0.12\\
\hline
\end{tabular}
\begin{tabular}{|p{.498\textwidth}|p{.132\textwidth}|p{.105\textwidth}|}
mean of absolute value of  five nightly averaged residuals ms$^{-1}$&0.32&0.11\\
\hline
\end{tabular}
\begin{tabular}{|p{.498\textwidth}|p{.132\textwidth}|p{.105\textwidth}|}
RMS of five nightly averaged residuals ms$^{-1}$&0.39&0.13\\
\hline
\end{tabular}
\caption{ Statistics per night on Standard deviation and Mean residual, both in ms$^{-1}$. Keeping the 200 exposures for night 25, the mean absolute residual (averaged over the 5 nights), and the standard deviation are computed in the two last lines}
\label{tab:residuals}
\end{table*}

The mean nightly residual (which can be qualified as a bias varying from night to night) is significantly nearer zero for the EPiCA method than for the official pipeline. More precisely, the mean of the absolute value of the nightly averaged residuals of the five nights 24 to 28 (we exclude the first night 22 which has only 10 exposures) is 0.11 ms$^{-1}$ for the EPiCA method instead of 0.32 ms$^{-1}$ for the Gaussian fit. Similarly, the RMS of the five nightly averaged residuals is 0.13 ms$^{-1}$ and 0.39 ms$^{-1}$ respectively for the two methods.
This is a measure of the quality of the fit to a 3 planets Keplerian model: a factor of 3 in favour of the EPiCA method.  In our mind it is a strong indication that this new method of applying CF1 to CCF$_{tot}$ is giving a result which is nearer the truth than the Gaussian fit. Admittedly, the factor 3 improvement of Kepler fit to a RV time series by using EPiCA method is tested on only one example, and is likely not representative of all RV time series collected so far.
It should also be kept in mind that such low values, 0.11 ms$^{-1}$, or even 0.32 ms$^{-1}$, are anyway  excellent, and reflects the high quality of ESPRESSO data, which is certainly even better after the upgrade of the blue detector cryostat in June 2022.

\begin{figure}
    \centering
    \includegraphics[width=0.5\textwidth]{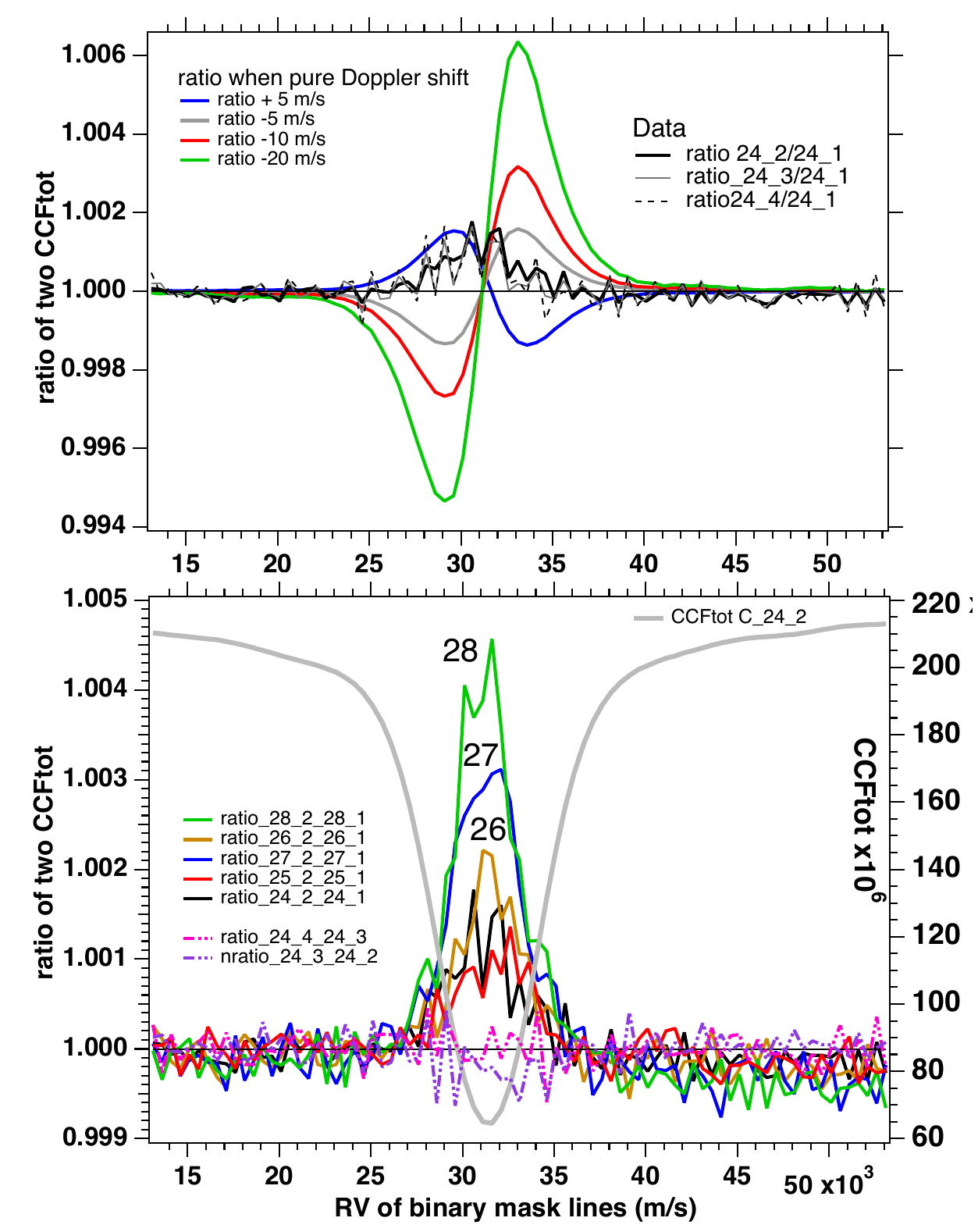}
    \caption{Model-data comparison using CCF ratios. Top: expected ratio of two CCFs with unchanged shape and various Doppler shifts (colored lines), and ratios using the four periods of night 24 (black lines). The same curve is observed for ratios of parts 2, 3 and 4 to part 1. It does not resemble the expected variation for a simple shift and indicates a change of shape during the first and second parts of the night. Bottom: Ratios of pairs formed by the first two consecutive stacked CCF$_{tot}$ obtained during the same night, for nights 24 to 28 (numbers indicated on the plot for 26 to 28). One CCF$_{tot}$ is drawn (grey line). There is a systematic {\it bump} centered on the CCF$_{tot}$, indicative of an increase of signal near the center during the first and second parts of each daily pair. The increase is larger for the latest nights. The ratio for the two last parts of night 24 is shown for comparison (pink line). There is no increase in this case, as sign of absence of shape variation during this period. }
    \label{fig:figp6_1_2}
\end{figure}

\begin{figure*}
   \centering
    \includegraphics[width=0.8\textwidth]{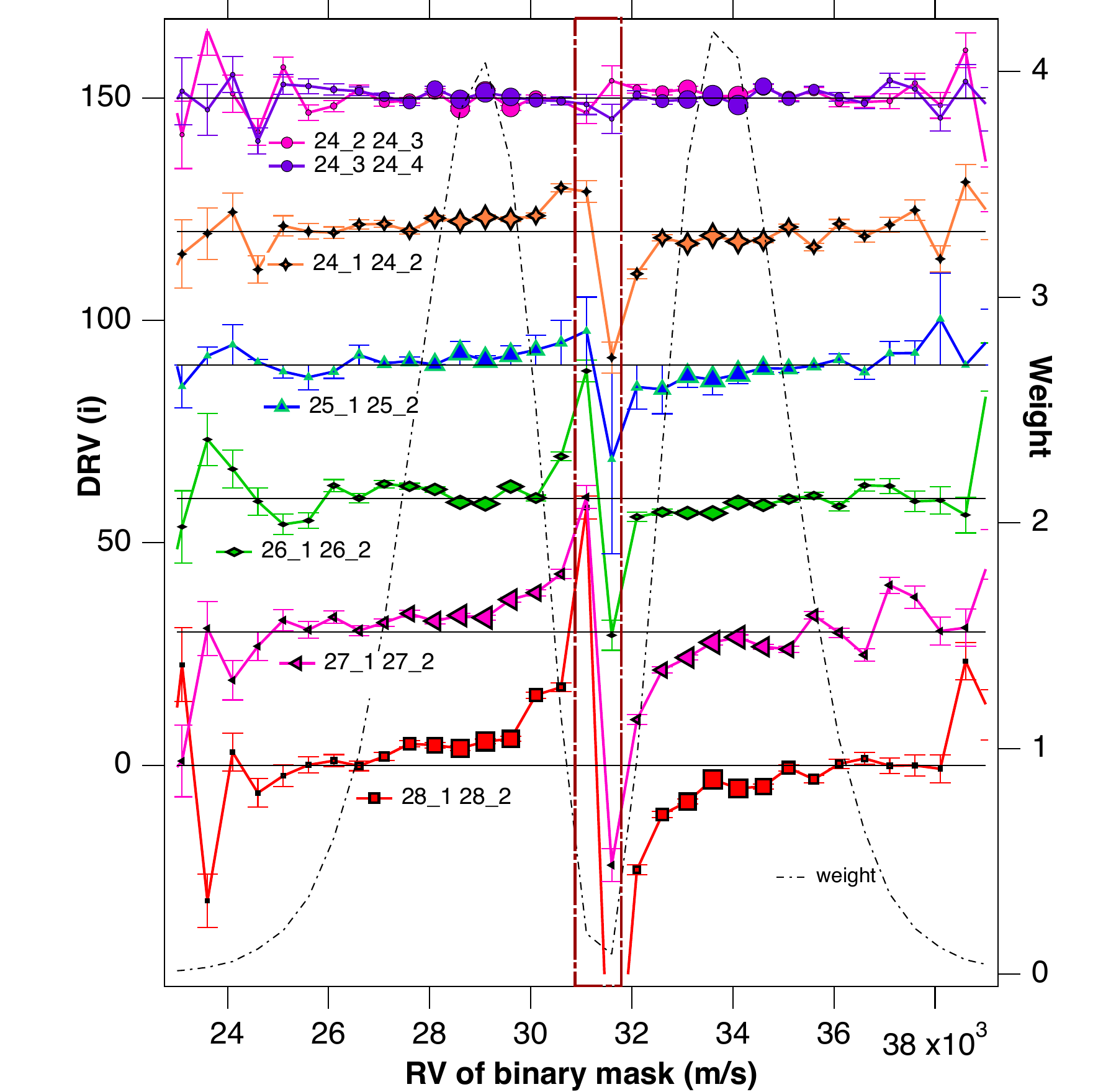}
    \caption{EPiCA profiles DRV(i) along the velocity grid for pairs of time periods (lines with markers). Offsets of 30 ms$^{-1}$ are applied between two consecutive profiles, except for the two top curves which are plotted together. The actual zero value is indicated by a black thin line. The size of the markers is increasing with the individual weight for each series. A typical weight function is also plotted (thin dot-dashed line). The two most central values suffer from additional uncertainties due to the limited velocity sampling and are delimited by a rectangle. (Top curves) For pairs made of the second to fourth parts of night 24, the profiles are flat within error bars, showing no significant change of shape: this is consistent to the constancy observed using ratios, pointing to a stabilization of the instrument.  For each of other pairs (lower curves) containing the first and second parts of a night, one can observe departures from a flat profile, and a similar shape with an amplitude increasing from night 24 to night 28. This typical shape of DRV(i) is induced by a decrease of contrast (see text), consistent with the same type of behaviour as the {\it bump} measured using CCF ratios (Fig. \ref{fig:figp6_1_2}). It could be due to a stellar process, or to an artefact.}
    \label{fig:figp6_newEPiCA1}
\end{figure*}

\section{EPiCA method and detection of changes in the spectral shape of stellar lines.}\label{variability}
As detailed above, the EPiCA method provides, for each point i of the radial velocity grid used in the building of the CCF$_{tot}$, an estimate of the change of radial velocity DRV(i), with an associated error bar, between two exposures at epochs E1 and E2. In previous sections we considered only the change of RV obtained by combining optimally the various values of DRV(i). However, it is also possible to examine the curve DRV(i) in detail, as a function of point i. If there is no change in the CCF$_{tot}$ shape between the two exposures, the DRV estimate should be constant across the RV grid. If it is not constant, then it means that the spectral lines shape have changed between the two epochs E1 and E2. Since CCF$_{tot}$ is an “image”, combination of all spectral lines in the stellar spectrum, this change must reflect at least a physical effect that is dominating in the average “image”, even if it is not present identically in all the spectral lines. 

In order to increase the SNR (up to about 3,000), we have stacked the CCF$_{tot}$ by 100 consecutive exposures whenever possible (taken over about 2 hours), and a little less otherwise. Night 24 contains 406 exposures, and is divided in 4 periods. There are 200 exposures for night 25, thus divided in two periods; night 26 has 174 exposures, divided in two parts; same for night 27 with 176 exposures and night 28 with 185 exposures. Hereafter we name for night XX and first (resp. second) series of exposures by C-XX-1 (resp. C-XX-2). Before any computation of a ratio, the two curves were normalized to each other, to account for photometric changes and to get a value around 1 at the extremities of the curves, corresponding to the shoulders of CCF$_{tot}$.

Changes in the spectral shapes can also be detected directly by dividing a CCF of one exposure  by the one of a different exposure. This can be done also using CCFs averaged over several exposures. In this section we have taken advantage of the unique series of short exposures to explore the application of EPiCA for this purpose and to compare with divisions. We explain why the diagnostic using EPiCA is easier to perform. 

Doing this study, we found a small change of the spectral shape during the first two hours of exposures (about 100 exposures, first quarter of night 24 and half of the total observation period for nights 25, 26, 27, 28), a change similar for all five nights, and diagnosed with both CCF ratios and with EPiCA. 
Because of this systematic, repetitive behavior and the similarity of shape variations for the five nights, we separated their study from the study of the other types of spectral shape variability.
The former effect is described in the first part of this section, and, in order to avoid the inclusion of its type of variability, we restricted the analyses presented in the second section to comparisons between either the two first parts, or the two second parts of the nights. 

 \subsection{Change of spectral shape between the first and subsequent periods of time within the same night} \label{shapes}

\subsubsection{Using CCF ratios} \label{shapes_night_ratios}

If there is a simple shift between two curves CCF1 and CCF2, their ratio should present a characteristic shape displayed as smooth coloured curves on figure \ref{fig:figp6_1_2} (top), the results of a simulation using the curve C-24-1 (first stack of night 24) and shifting it by -20, -10, -5, +5 ms$^{-1}$. By comparison, the computed ratio C-24-2/C-24-1 (after normalization) is characterized by a centered {\it bump}, quite different from any of the model curves. The ratio curves C-24-3/C-24-1 and C-24-4/C-24-1 are very similar to the ratio curve C-24-2/ C-24-1, showing that the shape of C-24-3 and C-24-4 are very similar to the shape of C-24-2 :  the variation producing the {\it bump} occurred only between the first and second parts of the night 24, and the shape did not change afterwards during this night 24. This is confirmed by the ratios between part 2 and 3, and also part 3 and 4 of night 24, displayed in Fig. \ref{fig:figp6_1_2} (bottom panel), which are constant within the fluctuation level observed for all ratios.   
Actually, we found out a similar central bump (and thus systematic behaviour) when making the ratios of second part to first part of the night C-XX-2/C-XX-1 for the other nights XX=24 to 28, as shown on figure \ref{fig:figp6_1_2} (bottom), with the central {\it bump} increasing from day to day. Such a centered bump in the ratios of CCF is the signature that the so-called contrast (relative depth) is decreasing with time between the two CCF. 
A change of contrast is often used classically as a stellar activity indicator. This is quite possibly the case here for HD40407. However, in the present case, because of the somewhat systematic behaviour of contrast change along each night of observation on a time scale of $\simeq$ two hours, we cannot completely exclude that this effect is connected to an artefact of so far unidentified origin.

\begin{figure}
    \centering
   \includegraphics[width=0.48\textwidth]{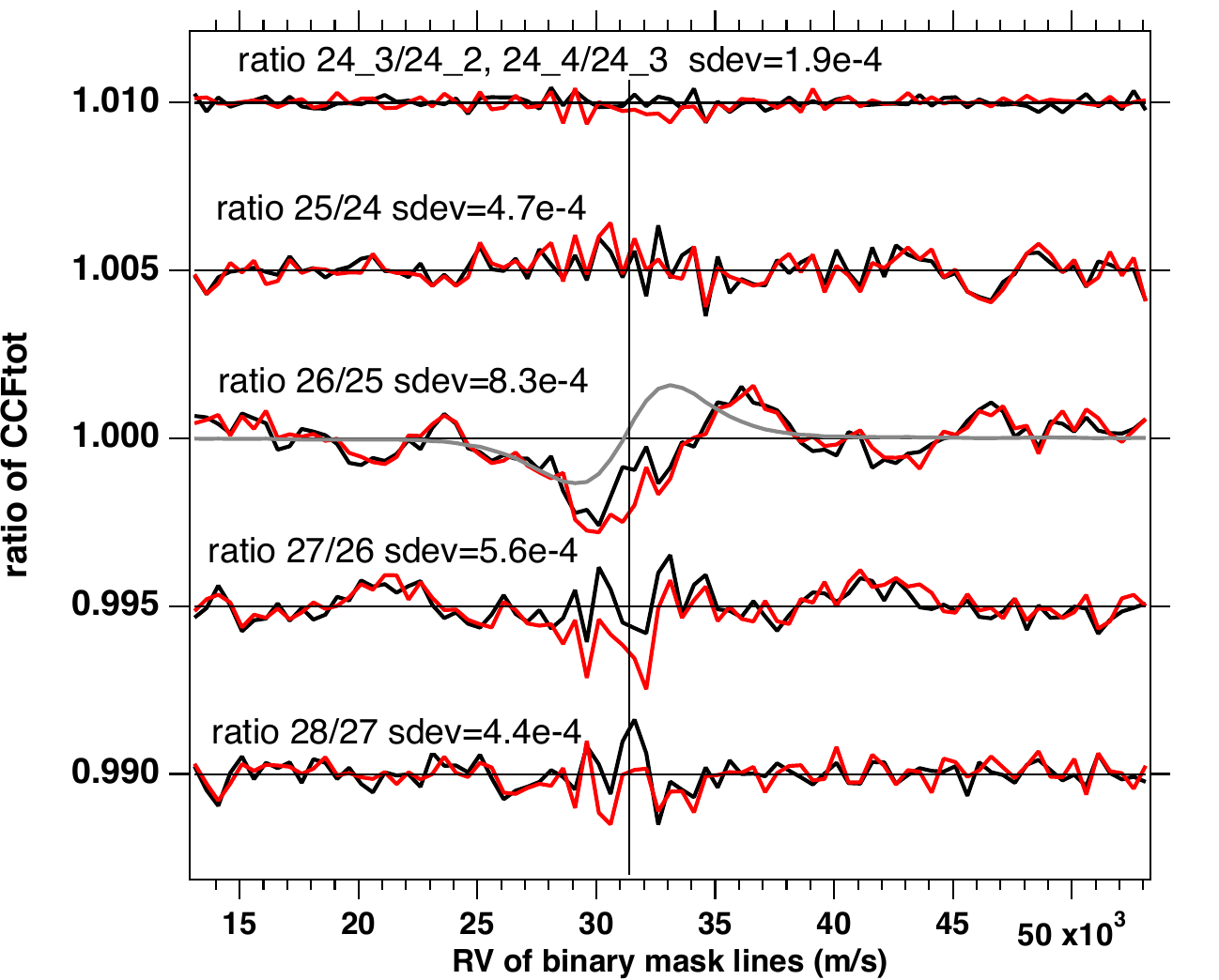}
    \caption{Ratios of pairs of CCF$_{tot}$ taken one day apart along the HD40307 campaign, as a function of RV(i).  There are two pairs for each couple of consecutive nights. The ratios are vertically displaced by multiples of 0.005 for each pair for clarity. The first and second pair of the night are respectively in red and in black. The standard deviation of the curve is indicated. The upper curves are two ratios of night 24, whose wiggles are small and likely representative of the noise level for all curves. There is marked departure from flatness for the ratio 26/25, as expected from the Kepler fit RV drop between nights 25 and 26. However, the observed ratio is very different from a model ratio in the case of a pure Doppler shift, as can be seen from the comparison with the superimposed model for a pure shift of – 5 ms$^{-1}$ (grey curve).  Therefore, this set of curves indicates directly a change of shape between the two nights.}
    \label{fig:fig_RATIO_PERIOD}
\end{figure}

\begin{figure*}
    \centering
   \includegraphics[width=0.6\textwidth]{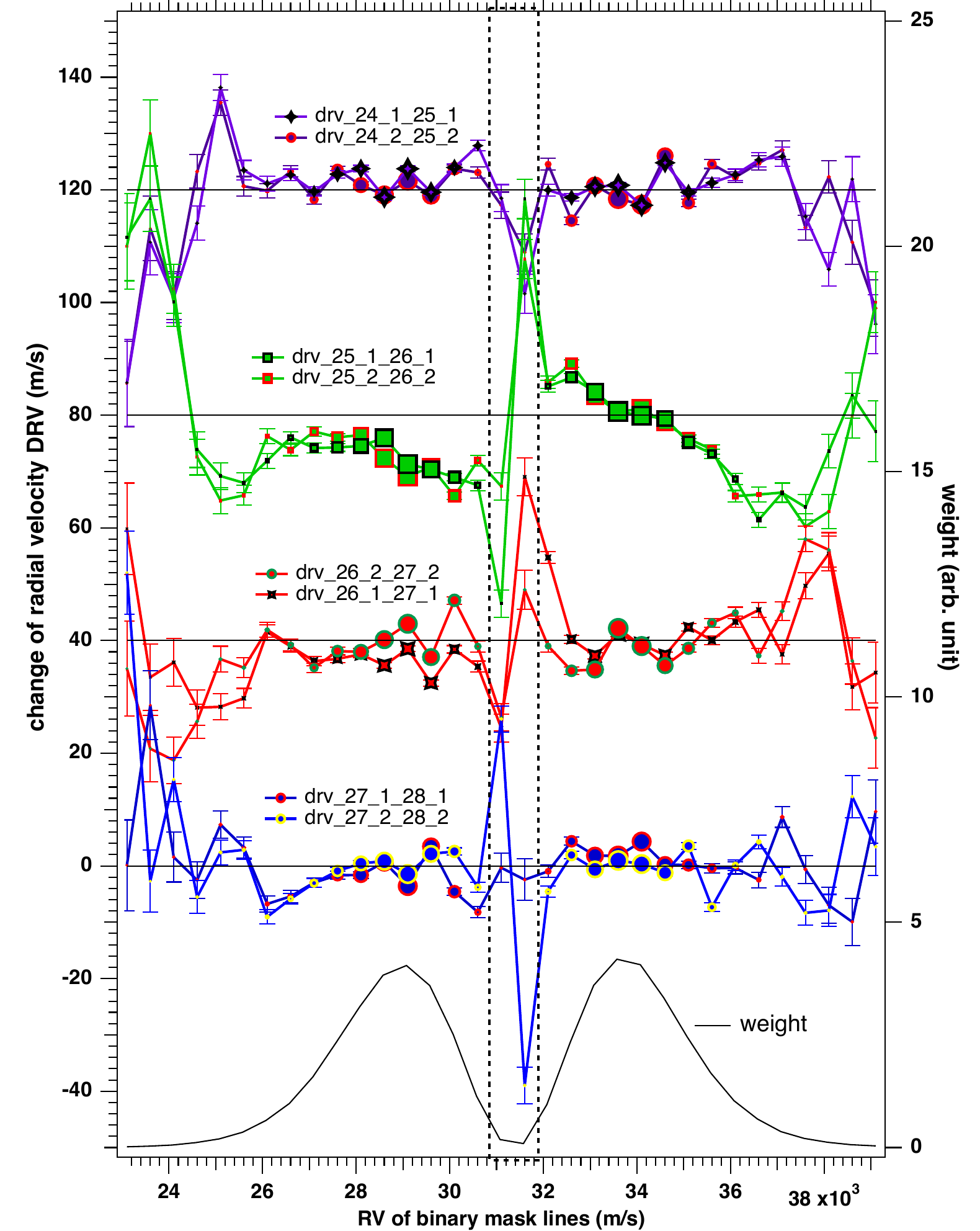}
    \caption{Same as figure \ref{fig:figp6_newEPiCA1}, for 4 pairs of stacks that are 1 one day apart over the 5 nights from night 24 to 28, using for each pair first and second parts of night. Offsets of 40 ms$^{-1}$ are applied between two pairs of these EPiCA profiles. The actual zero value is indicated by a black thin line. The DRV curve for the pair 25-26 shows a quite different pattern in comparison with the other pairs, with a strong difference between the blue side and the red side.}
    \label{fig:fig_EPiCA_PERIOD}
\end{figure*}

\subsubsection{Using EPiCA}\label{shapes_night_EPiCA}
In this subsection we examine how the change of contrast along a night revealed by the ratios of CCF of the previous section translates into the shape of DRV(i) curves derived from EPiCA. Fig. \ref{fig:figp6_newEPiCA1} displays several profiles along the velocity grid DRV(i) obtained with EPiCA for pairs of the same types of time periods as used in the previous analysis based on CCF ratios.  For this figure, the two members of each pair belong to the same night. Error bars are computed for each pair based on the second Connes formula (Equation 9) and reflect the weights associated with the slope of the CCF. A typical series of weights is superimposed.  The two most central values suffer from the division by the very small derivative (equation 3) and additional uncertainties due to BM coarse velocity sampling (500 ms$^{-1}$) and the corresponding area is marked by a rectangle. 

In case of a pure Doppler shift, i.e. in absence of a change of stellar line spectral shape, the DRV(i) curve should be flat. This is indeed the case (see top of figure \ref{fig:figp6_newEPiCA1}) for the two DRV(i) profiles for night 24 which are inter-comparing the second and third, then third and fourth parts. They are flat, within the error bars: there is no change of contrast, as revealed by the ratios of curves on figure  \ref{fig:figp6_1_2} (bottom) ; the instrument  has stabilized. 
On the contrary, the other DRV(i) curves of figure \ref{fig:figp6_newEPiCA1} depart significantly from a flat , constant, value. For all pairs of periods containing the first and second parts of the night, the profiles exhibit departures from zero on both sides of the central velocity. Moreover, the amplitude of these features increases from night 24 to night 28, forming extended {\it spikes}. Also, there is a difference of average level between the two sides of the curve DRV(i): it is positive on the blue side ($\leq$31.4 kms$^{-1}$), and negative on the red side. This is consistent with what is expected form the Connes algorithm applied to this particular case of change of shape, i.e. when the CCF$_{tot}$ becomes shallower between stack 1 and stack 2 (decrease of contrast); the increase of intensity of the blue side is interpreted by the algorithm as a red shift, while the increase of intensity of the red side is interpreted as a blue shift. This case is similar to one of the model exercises performed by \cite{Cretignier20}, comparing the Gaussian fit of a spectral line with the algorithm of Connes, as we do here (their figure B3). Namely, it is the third row of their figure B3, where they model a decrease of depth (or contrast) induced by temperature-sensitive lines. All in all, both approaches (ratio curves and DRV(i) curves) are consistent, pointing to a small decrease of the contrast of CCF along each night.

\subsection{Change of spectral shape between two consecutive nights}\label{shapes_period}

Whatever the cause of the observed change in contrast between the first part of the night ($\simeq$2 hours or 100 exposures)
 and the rest of the same night, we have also studied the change of shape from one day to the next by comparing two stacked CCF$_{tot}$ taken similarly along each of the two nights, i.e. using pairs of type C-25-1/C-24-1, C-25-2/C-24-2, C-26-1/C-25-1, C-26-2/C-25-2, and so on.

\subsubsection{Using CCF ratios}\label{shapes_period_ratios}

The CCF$_{tot}$ ratios for two consecutive nights  are displayed on figure \ref{fig:fig_RATIO_PERIOD}. There are two curves for each couple of days, one for the pair of first CCF$_{tot}$ of the nights, one for the pair of second CCF$_{tot}$ of the nights. The two curves for pairs of the same two nights are very similar, even in their detailed structure, in spite of the fact that they are built from different, and fully independent data sets, bringing some confidence in their credibility. The curves are different from day to day. The ratios are close to constant, except for the pairs 25/26 which are characterized by positive and negative values far above the noise. This large change between the two nights (quantified by the standard deviation over 81 points of the ratio)  happens at the time of the largest RV change ($\simeq$ - 4 ms$^{-1}$) predicted by the planetary model (figure \ref{fig:fig5}), but, of interest here, also at the time of the abrupt change (by $\simeq$ - 1.3 ms$^{-1}$ , figure \ref{fig:fig3_4}) of the difference between the two time series, Gaussian fit to CCF$_{tot}$ and EPiCA method. Importantly,  the change of shape of the ratio observed from day 25 to day 26 is quite different from the model prediction for a simple shift (see a model for a shift of – 5 ms$^{-1}$ superimposed for comparison in Fig. \ref{fig:fig_RATIO_PERIOD}). This observed departure from  a pure Doppler shift clearly indicates a significant change of spectral shape between night 25 and night 26, 
quantified by sdev= 8.3 10$^{-4}$. The other night ratios show smaller changes, quantified (in decreasing order) by sdev=5.6 10$^{-4}$ (27/26), 4.7 10$^{-4}$ (25/24), 4.4 10$^{-4}$ (28/27). The ratios for pairs of the same night 24, 24-4/24-3 and 24-3/24-2 have a sdev= 1.9 10$^{-4}$, representative of the noise level for such ratios. The two red and black curves of 27/26, and also the ones of 28/27 ratios are different from each other near the center, because of the contrast artefact which is more severe for nights 27 and 28 than for the previous nights, as shown on Figure \ref{fig:figp6_1_2} (bottom).

\subsubsection{Using EPiCA}\label{shapes_period_EPiCA}

Similarly to the previous section using the CCF ratios, we searched for changes of spectral shape from one night to the other using EPiCA and the same parts of the night. Fig. \ref{fig:fig_EPiCA_PERIOD} shows the results separately for each pair of consecutive nights. The two series of profiles, using the first or the second time periods of the two consecutive nights, are very similar, despite being derived from independent data sets, in the same way CCF ratios were similar. The DRV curves for the pairs of nights  24-25, 26-27 and 27-28 show somewhat constant values near 0 for both the blue and red sides (still restricting to the data points with low error). This is consistent with a pure, very small Doppler shift. In the case of the two DRV curves between night 25 and 26, their shape is very different from the other DRV curves, and, clearly, these DRV curves of the pair 25-26 stand alone. One would expect a flat profile at about -4 ms$^{-1}$ in both the blue and red side, according to the planetary model. However, the two sides of the curves are very different from each other. The blue side is negative, which is consistent with the planetary motion between day 25 and 26 (figure \ref{fig:fig5}), but decreases strongly below -4 m.s$^{-1}$ (note that we discard from discussion the two spikes near the center). The red side is also not constant, and decreases but is globally positive. Such behavior  is showing that the shape of the CCF$_{tot}$ profile has changed substantially between night 25 and night 26. This is totally in line with the change of shape characterized by the ratios of CCF$_{tot}$ displayed on figure \ref{fig:fig_RATIO_PERIOD}, where also the pairs of ratios between nights 25 and 26 stand alone among one day apart ratios, with a pronounced change of shape, not symmetrical with the center of the line.

\subsection{Conclusion about the line shape variations and the two methods of detection}\label{shapes_ratio_EPiCA}

In summary, we found that the DRV(i), the Doppler velocity shift between two CCF$_{tot}$ is sometimes very far from being constant along the RV grid points i. This is the proof that the spectral lines are changing their shape, in addition to any Doppler shift, at a time scale of $\simeq$2 ~hours or a little shorter. With the help of the ratio curves, we detected a somewhat systematic effect: a central “bump” builds up between the first stack of the night, and the second stack, along each of the 5 nights, possibly of stellar origin or resulting from an artefact. Whatever its origin is, we mitigate this effect by comparing CCF$_{tot}$  taken at the same relative time during the nightly operation and one day apart. Doing so, we detected a significant variation of shape, between night 25 and night 26, exactly at the time of the abrupt change  of the difference between the RV time series deduced from the Gaussian fit and the EPiCA method (decrease by $\simeq$ 1.3 ms$^{-1}$).  This suggests that the change of shape modifies differently the results of the Gaussian fit and the shift finding algorithm. We assign this change of shape to the stellar activity.

For both types of detection, it is interesting to compare with the lineshape simulations of \cite{Cretignier20} displayed in their figure B3. First, the authors conclude that, adding a symmetric perturbation to a symmetric spectral line, both methods, Gaussian fit and Connes's formalism (equivalent to EPiCA) are robust, and the retrieved RV does not change. On the contrary, if a symmetric perturbation to a non-symmetric line (which happens at least whenever there is a blend of a smaller line in one side of the line), then both algorithms are retrieving a somewhat biased change of RV. The “bump” centered described in \ref{shapes_night_ratios} resembles a case of a symmetric perturbation, such as a change of contrast. It is very likely that a fraction of lines from the binary mask has blends. But since secondary line blends are distributed at random, the CCF$_{tot}$ will be much more symmetric than a single blended line, and it is likely that both methods are robust to this artefact, yielding small biases if any. 
The case of the addition of a non-symmetric perturbation is not shown in their exercise, but, from their exercises of addition of a symmetric perturbation,  we certainly may expect that a biased DRV will be retrieved for both methods. Actually, the (largest) change of shape observed from night 25 to night 26 is strongly asymmetric (figure \ref{fig:fig_EPiCA_PERIOD}), and therefore we may expect that the DRV retrieved by the two methods will both be biased erroneously away form the true dynamical change of RV due to planets. Because this peculiar behaviour between night 25 and 26 corresponds exactly to the abrupt change of the difference between the two time series (drop by $\simeq$ 1.3 ms$^{-1}$), the classical Gaussian fit and the EPiCA method, as displayed on figure \ref{fig:fig3_4}, this difference in the time series of $\simeq$ 1.3 ms$^{-1}$ is certainly the difference of biases between the two methods. 

Which method provides the smaller bias? Extensive simulations are beyond the scope of this paper. Whatever the answer, we may emphasize that EPiCA is a more direct and easier way to detect shape variations. As a matter of fact,  departures from flat DRV(i) curves are an easy diagnostic of shape variation, they cannot be confused with effects of a pure Doppler shift, and there is no need for a comparison with a model, as in the case of CCF ratios. We also argue that the EPiCA method is easier and has more flexibility in the analysis of departures from a single Doppler shift between two epochs than the comparison of CCF bisectors.  By essence, the bisector is making a link between both sides of the spectral lines (or the CCF image), while the Connes algorithm is applied independently to each point of the line (or the CCF image). The bisector says nothing about the length of segment joining the two sides of the line: the center of the segment may not move, while the length of the segment may change without being noticed by the bisector analysis. One may oppose that in this case there is no bias in RV, but symmetric variations like those detected at the beginning of the observing periods would not be detected.

The variation of stellar line shape from night 25 to night 26 is likely a result of stochastic changes of the convective granulation blue shift (GBS) averaged over the stellar disc \citep{Meunier2020,Meunier2021}. Model results of these authors indicate an amplitude of RV Doppler shift fluctuations of $\simeq$0.2 m.s$^{-1}$ for a star of K4 type (similar to HD4030, type K2.5), when granulation only is considered with time scales of 15 - 50 minutes, and up to $\simeq$1 m.s$^{-1}$ for super-granulation changes (on a $\simeq$one day time scale). Such a super-granulation change could explain the important departure from constancy of the DRV(i) curve, and the corresponding change by $\simeq$1.3 m.s$^{-1}$ between the difference of RV retrieved either from the Gaussian fit or from EPiCA method, and remaining during the following nights 27 and 28.

\section{Summary and discussion}
\label{sec:conclusion}

The main reason is that the spectral line shapes are not symmetric, while the Gaussian fit is symmetric, which produces a bias when finding the center. And this bias is not constant in time, because the spectral line shapes are changing with stellar activity, as revealed by bisector analysis. 

In this paper we have described and tested a new way to estimate the change of radial velocity DRV between two epochs from the comparison of the two stellar spectra. This algorithm is based on the first formula of Pierre Connes, which aim is exactly this, and was shown to be the (mathematically) optimal way to retrieve DRV. Instead of applying the shift finding algorithm to two spectra, as it was originally designed, or to individual spectral lines, as in some line-by-line techniques, it is applied here to the weighted sum of all CCFs obtained separately for each order, CCF$_{tot}$, which represents an average "image" of all spectral stellar features. If the computation of CCF$_{tot}$ is done on a wavelength scale associated with the solar-system barycentric frame, i.e., using the corresponding barycentric velocity grid, the reflex motions induced by planets are small and one can stop at the first order term of the Taylor expansion involved in the formalism of equation 2. This new EPiCA method is simple to implement and appropriate when pipeline data products of the spectrograph include indeed CCF$_{tot}$ in the solar system barycentric frame. It combines the advantage of the shift finding algorithm and the robustness of CCF construction.\\
To test the algorithm, we have used a series of spectra acquired during a one-week observing campaign on star HD 40307, which is known to host at least 3 planets and we compared the two RV time series obtained from the CCF$_{tot}$ profiles by means of a Gaussian fit on the one hand, and our new EPiCA algorithm on the other hand. In order to compare quantitatively the two methods, two different "quality" indicators of the retrieved RV time series were computed. \\
First, a single-night series was examined. The RV series were fitted to a linear relationship representing the first order Doppler variation under the influence of planets, but also any RV change that could be due to stellar activity. The first order approach is justified by the short duration for the analysis. The dispersion of the residuals to the linear fit, which is a measure of the mean error, was found to be $\sigma$=1.03 and 0.83 ms$^{-1}$ respectively for the pipeline and the EPiCA algorithm, a significant 20\% improvement of the precision on RV changes. We interpret this decrease of the RV dispersion as an evidence for the absence of strong variations of the stellar line shapes during this time interval, at least for the absence of variations strong enough to cancel the benefit of the application of the Connes shift finding algorithm which makes an optimal use of all the spectral information but requires unchanged profiles.\\
In a second test, we made use of the whole series obtained over the campaign, and the computed RVs were fitted to a 3 planets Keplerian model. The periods and reflex-amplitudes were fixed to the previously determined values, but the phases were adjusted. Based on the nightly-averaged residuals to the respective best fits (figure \ref{fig:fig11}), we computed a second quality indicator, defined as the mean over 5 nights of the absolute value of nightly mean residual. This average bias was found to decrease significantly with the new algorithm. We found 0.32 and 0.11 ms$^{-1}$ respectively for the pipeline and the EPiCA algorithm, i.e. a factor of 3 of decrease. Inasmuch as a 3 planets Keplerian best-fit is "the dynamical truth", not affected by stellar activity, it appears that the EPiCA algorithm (at least in the HD 40307 study case) is nearer the truth than the Gaussian fit. As a side result, we found a somewhat intriguing new value of the orbital period of planet b, averaged over 10 yr from 2008 to 2018. \\ 
The main difference between the two RV series, at the origin of the differences in the model fitting, is associated with the significant increase of the difference between the two determinations between night 25 and night 26, by about 1.3 ms$^{-1}$. We come back below to this phenomenon, associated with a quite significant line shape variation. While, as mentioned above for the first test, it is easy to explain the benefit of the EPiCA method in the case of negligible changes of the line shapes, it is not totally obvious why the EPiCA method should provide a more accurate estimate of the changes of dynamical RV than the Gaussian fit method for long periods of time and strong line shapes variations, which seems to be the case based on the results of our second test. Admittedly, the comparative tests of the two methods shown in this paper are limited to only one example of one stellar system. However, our results are encouraging and show that such a method deserves to be explored further. It might be worth to get a few hours, high cadence, time series at least once for each star monitored for the search of exoplanets. It should allow to estimate whether or not the EPiCA method provides a better accuracy in the (likely) case of small line shape variations. A study similar to our second test, on the other hand, on longer duration for a star without planets would be very useful too. 
It would give some insight on the relative sensitivity of the two methods to line shape variations specific of the observed object. \\

About the abrupt change of the offset between the two time-series between night 25 and night 26 (figure \ref{fig:fig3_4}), we assign this significant change of offset by the combination of two effects:
\begin{itemize}
    \item stellar processes (most likely supergranulation blue shift, because of the about one day time scale) are present and modify substantially the shape of the spectral lines and the resulting CCF$_{tot}$ between the two nights.
    \item the two algorithms are not sensitive in the same way to those changes of spectral shapes. 
\end{itemize}

 If the results of the second test are interpreted as a smaller sensitivity of the EPiCA algorithm to stellar changes, this smaller sensitivity is very likely linked to the basic difference of the two algorithms. The standard Gaussian fit to the CCF$_{tot}$ assumes that the Gaussian is symmetric, while the EPiCA method reveals that the change of shape between two epochs may be strongly asymmetric (see figures  \ref{fig:fig_RATIO_PERIOD} and \ref{fig:fig_EPiCA_PERIOD}). Therefore, fitting always with a symmetric Gaussian a curve which is not symmetric, and which is varying in its asymmetry, will introduce a bias in RV, variable from epoch to epoch. 
 

An amplitude of 1.3 ms$^{-1}$ is significantly larger than theoretical estimates of the RV RMS of granulation blue shift (GBS), based on stellar atmosphere models: $\simeq$ 0.1 ms$^{-1}$ \citep{Cegla19}; 0.26-0.30 ms$^{-1}$ for a K1 star \citep{Meunier23}. This is why we believe rather on supergranulation variability. On the other hand, this observed value of change by $\simeq$ 1.3 ms$^{-1}$ is quite compatible with faster RV variations observed on star HD 88595 (F7V) as reported in \cite{Sulis23} that they assign to GBS. As a matter of fact, even after binning over 2.5 hours their RV time series (obtained with Gaussian fit to CCF), their resulting RV RMS is still $\simeq$ 1.5 ms$^{-1}$.\\

These results imply that the measurement of the offset between the Gaussian fit RV (the pipeline RV) and the shift finding algorithm RV provide good indicators of significant stellar line shape variability, and it motivated us to pursue in this direction. Instead of only extracting RV for each exposure, we additionally tested an extension of the EPiCA method to the analysis of the EPiCA profiles along the velocity grid, with the goal of characterizing  the stellar line shape variations. Applying the shift finding algorithm between two CCFs allows to determine a change of RV (a shift) DRVi, for each grid point i of the radial velocity.  If there is only a simple shift between the epochs of CCF1 and CCF2, the curve DRV(i) should be constant with i. This is what we implicitly assume when we average all DRV(i) values to extract RV. On the contrary, any departure from constancy is a sign of line shape change, and consequently of stellar activity. 

The detailed study of the DRV(i) profiles provided several results. Firstly, we detected a somewhat systematic variation of the profile every night, occurring during the first two hours of measurement. The variation is weak and symmetric, and, \textbf{regardless of} its origin, it does not affect the RV determination. In order to avoid any interference with the study of the DRVi profiles on the longer term, we limited the study of the day-to-day variability to the inter-comparison of the same periods of the observing run, namely using the first (resp. second) 100 exposures of night N with the first (resp. second) 100 exposures of the following night. We found significant departures from profile constancy between the night 25-night 26 pair, and constancy for the other pairs. The non constant profile corresponds  exactly to the same interval for which we have the abrupt variation of the difference between the RVs from the Gaussian fit and from EPiCA. It is a second indication that a stellar line shape change has occurred. Here, however, we have more information on the type of variability that happened.
In this respect, it is useful to compare with the simulations of DRV(i) done for the LBL analysis by \cite{Cretignier20}. The shapes of the DRVi curves give an idea of the types of distortions the line shape suffered from.

In conclusion, the application of EPiCA to the merged CCFs may be quite useful in several ways. It can provide on-the-fly results as soon as two spectra are collected, along with the pipeline results. It may produce a smaller dispersion of the RV series compared to the Gaussian fit, and, independently, changes in the offset with the Gaussian fit RVs are a warning for line shape variations. Although it can not compete on the long term with LBL analyses, the use of CCF$_{tot}$ yields a much higher signal-to-noise ratio (SNR) than when the shift finding algorithm is applied on a line-by line basis, allowing the detection of global line shape change and stellar activity. It should be recognized however that our method would also benefit from a median template spectrum and better defined derivatives of CCF$_{tot}$, because the noise on DRV(i) would be substantially reduced. Finally, the EPiCA method is rather easy to implement, since CCF$_{tot}$ is already in the ESO data products for ESPRESSO, and the Python codes are provided to the user’s community (see explanations and information in Appendix A). \\

\begin{acknowledgements}
We thank our referee for his/her very constructive and detailed comments. We deeply thank Nadège Meunier for very detailed discussions about the role of convective granulation blueshift as an important perturbation to radial velocity measurements. We wish to thank Willy Benz suggesting the implications of a possible long term change of the period of HD40307 b. A.I. acknowledges the support of a Vernadski Scholarship for PhD students, sponsored by the French Government and the Ministry of Science and Higher Education of the Russian Federation under the grant 075-15-2020-780 (N13.1902.21.0039). 
\end{acknowledgements}

\bibliographystyle{aa}
\bibliography{biblio}

\begin{appendix}
\onecolumn
\section{Where to find ESPRESSO CCF and how to use the code}
With the new service of the 
\href{http://archive.eso.org/eso/eso_archive_main.html}{ESO archive} 
it is possible to obtain not only the raw data, but the ancillary data as well. For ESPRESSO at VLT, one of the type of ancillary data is the ANCILLARY.CCF - 2D array containing the cross correlation function of fibre A with a stellar template spectrum, computed per Echelle order. Each row represents one order, and the X direction is the sampling in velocity space.\\
The information about the velocity space (or grid) can be retrieved through the fits header, "HIERARCH ESO RV STEP" stands for the step in kms$^{-1}$, and "HIERARCH ESO RV START" is the first velocity value in kms$^{-1}$, the number of points at the grid is the same as the length of the CCF array.\\
The ANCILLARY.CCF has 171 orders, while ESPRESSO has only 170 orders, the last column of ANCILLARY.CCF is the result of adding all the single order CCFs together. In the current version of code the computation is provided using this column only, but it is planned to extend the code to provide more flexibility and allow users to choose which orders to consider.\\
In the current configuration the code works only with data provided in the fits format of ESPRESSO CCF files, in the future an extension for more custom formats (numpy array, text data) is being considered. To get started, one needs to specify the path to the reference CCF and the path to the other CCFs (in case of other CCF it is possible to provide path to the fits file if only one CCF is considered, text file with paths or just put the path to the directory if an ensemble of CCFs are considered). It is important to mention that the current version of the code does not check the date of observations and makes a loop on files in accordance to their position in the list/directory: an initial correct compilation list is important to get correct results.
The velocity grid is computed by the code itself on the basis of the fits header information.
The result is saved as numpy array.
All paths should be provided with '.txt' file. Examples of usage and documentation of further changes can be found at the GitHub repository.\\ 
Specific questions about the code should be addressed to  anastasia.ivanova@cosmos.ru.

%

\section{Putative change of period of planet b between 2008 and 2018.}

\begin{figure*}
    \centering
 \includegraphics[width=0.48\textwidth]{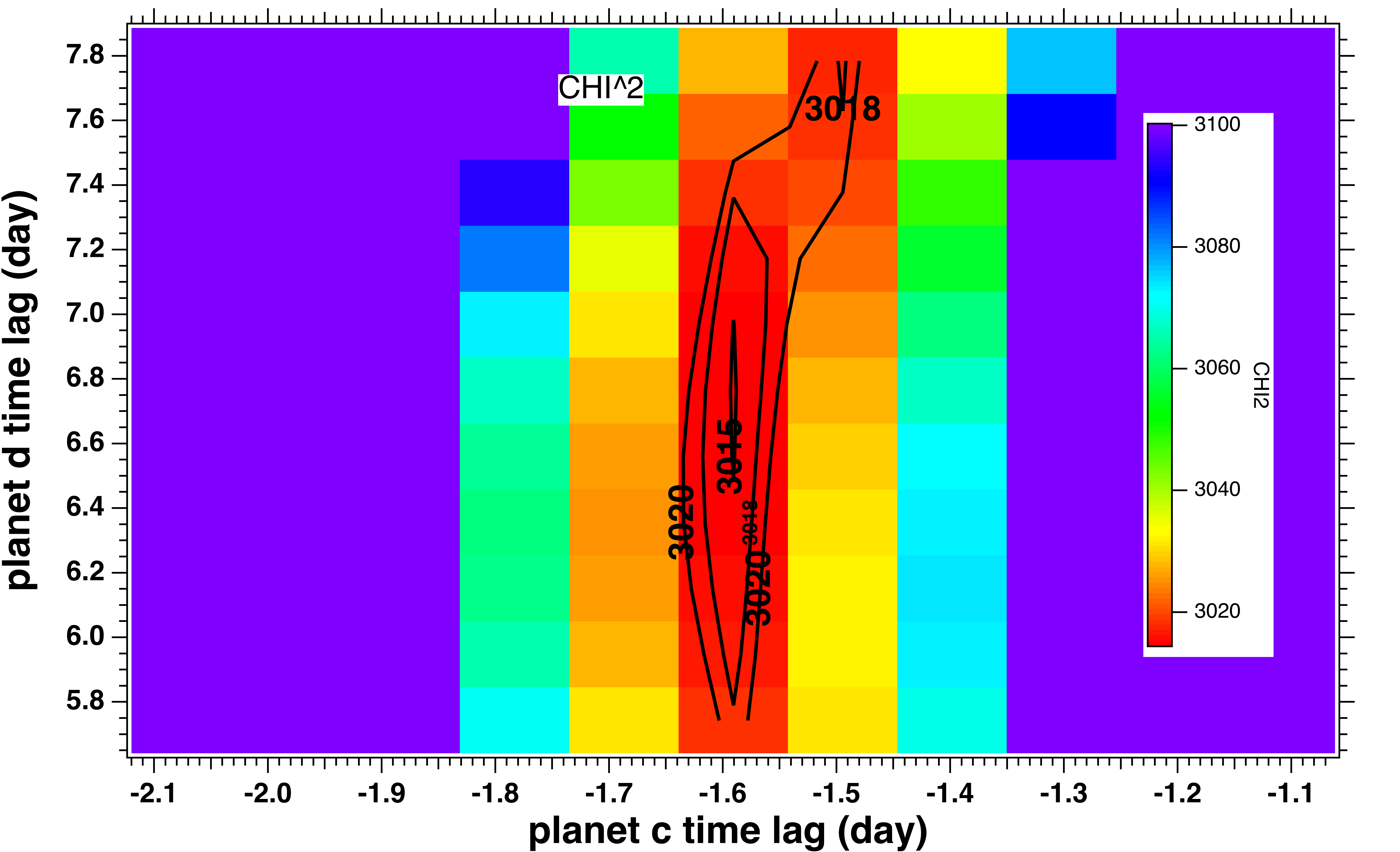}
\includegraphics[width=0.48\textwidth]{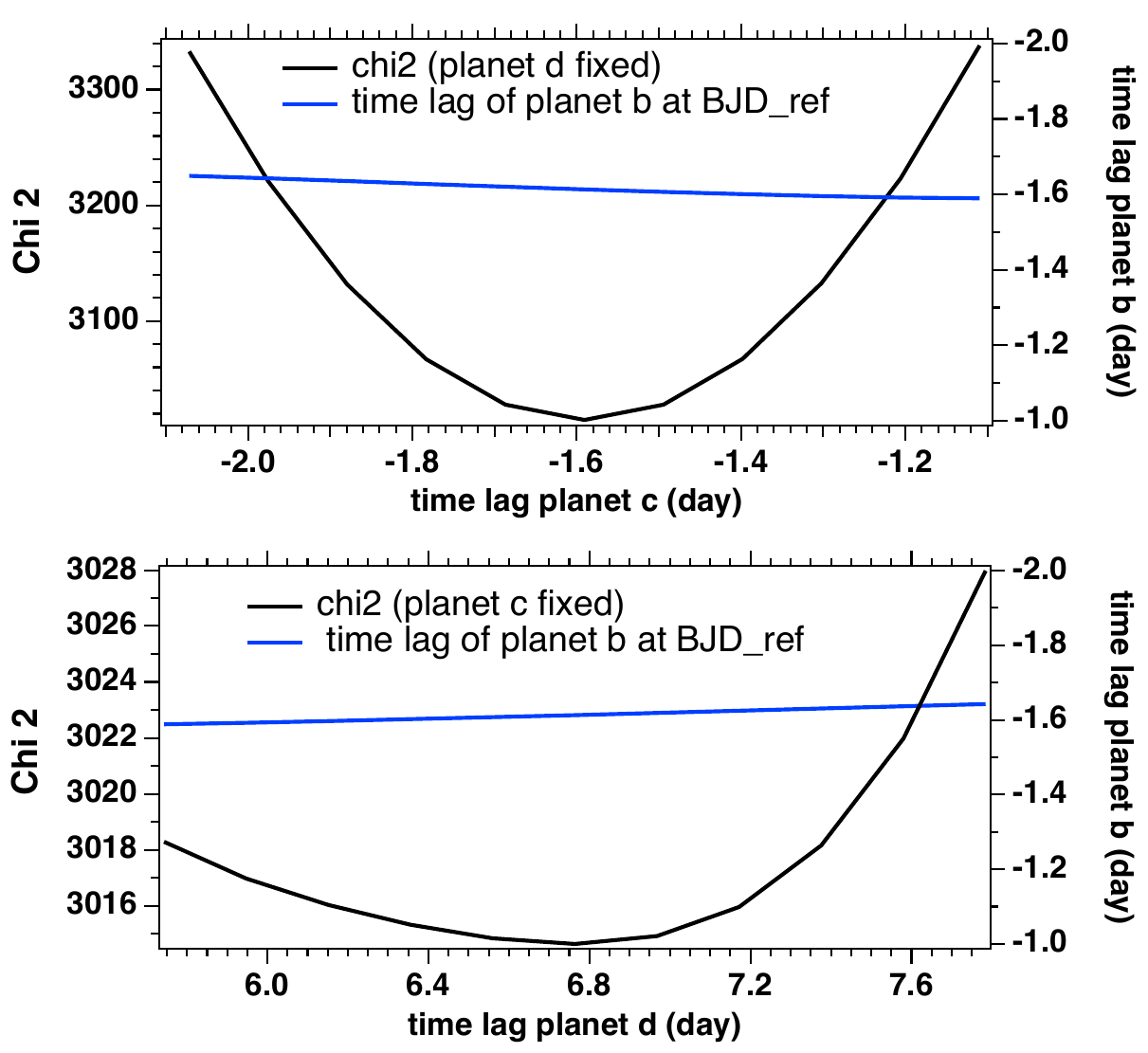}
\caption{Left: Iso-contours of Chi$^{2}$ when fitting the EPiCA time series with a 3 planets model. For each couple of planet c and d time lag with BJD$_{ref}$, a best fit of the time lag of planet b is computed, as well as the overall Chi$^{2}$ on the 1151 points. The minimum Chi$^{2}$ is at 3014, the average Chi$^{2}$ per point is 2.6 (ms$^{-1}$)$^{2}$. Top Right: Horizontal cross-section of left image, with the variation of the time lag of planet b, and Chi$^{2}$, as a function of the time lag of planet c, when the time lag of planet d is fixed to its overall best fit value. Bottom right: same as top but the roles of planet d and c are inverted. For both panels, the variation of the time lag of planet b is very small, showing an absence of coupling between the phases (or time lags) of the 3 planets.}
    \label{fig:fig_B1B2}
\end{figure*}

Here we discuss further the intriguing result of an apparent  change of period of planet b, between 2008 and 2018. Looking at figures \ref{fig:fig6} and \ref{fig:fig7} where the RV data are plotted with the 3-planets best-fit, it is rather clear that the observed RV drop from night 25 to night 26 by $\simeq$ 4 ms$^{-1}$ cannot be entirely due to planet c or planet d, because of their longer periods, and is mostly due to planet b with the shortest period (4.3114 d). Therefore, this sudden drop constrains very much the exact phase of the planet b sine wave, on which is based the finding that its average period has changed between 2010 and 2018. Quantitatively, we explored a potential coupling between the phases of the three planets with the following exercise, applied to the EPiCA RV time series.   
For a 2D ensemble of 11x11 possible phases at BJD${ref}$ of planets c and d, the 3 planets best fit was run, keeping fixed the phases of planets c and d, yielding the best fit phase of planet b and Chi$^{2}$. It yields a 2D image of Chi$^{2}$ displayed on figure \ref{fig:fig_B1B2}, showing some iso-contours of Chi$^{2}$. Two 1D sections of this image are also plotted on figure \ref{fig:fig_B1B2}, showing both the Chi$^{2}$ values and the phase of planet b (expressed in days, time lag= period x phase). On the top panel, the phase (expressed in days, time lag) of planet d is kept fixed at its 3 planets best-fit value, as a function of the time lag of planet c. Similarly, on bottom panel, the same curves of Chi$^{2}$ and time lag of planet b are plotted as a function of time lag of planet d when the time lag of planet c is fixed at its 3 planets best fit. While the Chi$^{2}$ increases significantly away from its 3 planets best fit, the time lag of planet b does not change very much, at -1.61 $\pm$ 0.03 d: it is very weakly coupled to the exact values of phases of planet c and d.Therefore, we may consider the retrieved phase at BJD$_{ref}$ of planet b as a robust result, almost independent of the phases of the two other planets, within the 2D area of the Chi$^{2}$=3018 contours. Consequently, the determination of T2b, the start time of the sine wave just before BJD$_{ref}$, is also robust at T2b = BJD$_{ref}$ -(4.3114-1.61) $\pm$ 0.03 d, as well as the average period between 2008 and 2018 as determined in section 5.4.2. Including in the best fit process planet f at 51.6 d period does not change the best fit values of the phases of other planets. 

Changes of periods due to gravitational interactions between planets have been detected, particularly by monitoring the Transit Time Variations (TTV) of transiting planets. This allows to determine the mass of the perturbing planet, as discussed for instance in \cite{Nesvorny08}. Actually, the period of the perturbed planet (time between two successive transits) changes periodically, typically by $\simeq$ 200 s, and with a period of typically $\simeq$200 days. 
In the transit geometry, the measurement of the time between two transits can be determined very accurately (within seconds), and changes of period detected easily. On the contrary, in a system with no transit and monitored with the RV measurements, such changes of period would be much more difficult to detect. Consequently, it cannot be excluded that the HD40307b planet is indeed in this typical TTV configuration.
The system around HD40307 was observed from 2003 to 2014 (only 7 observations prior to 2006), a time span of 10.4 yr \citep{Diaz2016}, before the 2018 observations used here. If planet HD40307 b were influenced gravitationally with a periodic change of period, then the reported period at 4.3114$\pm$0.0002 days is the mean period. The reported uncertainty on the period of 0.0002 d, or about 20 s, does not preclude at all a periodic change of its orbital period, even by 200 s typical of TTVs. On the other hand, the comparison of zero crossing times between epochs 2008 (T$_{0b}$) and 2018 (T$_{2b}$) yields a mean period averaged over 10 years, which is different from the first one by 0.0015 d, or $\simeq$ 130 s. Therefore, the change of period pattern found between 2008 and 2018 is quite different from the typical TTV pattern of $\simeq$200 days periodic change of the period by $\simeq$200 s. In addition, if the period determined over 2003-2014 had been the same as the one determined between 2008 and 2018, the difference of periods of 0.0015 day would have accumulated over the 1219 orbits of planet b up to a time lag of 1.828 day, or slightly less than one half of orbital period. It is very doubtful that such a discrepancy would have gone unnoticed by the various analysis of the period 2003-2014 time span \citep{Diaz2016,Coffinet2019}, supporting the analysis reported here. It sounds more like a longer time scale, secular change of the period that could be related to an eccentric planet, a very interesting subject indeed but out of the scope of the present paper, which deals with RV retrieval methods. Both methods yield similar results on this topic. A detailed analysis of the system stability is beyond the scope of the current work. Other scientists are encouraged to examine independently  the 2018 observations to verify the phase and period results presented here, even taking the standard official RV results contained in the data product. Both time series are available at the GitHub repository.

\begin{table*}[h!]
\begin{tabular}{ |c|c|c|} 
  \hline
&Planet b pipeline&Planet b CF1\\
\hline
   Amplitude and error [ms$^{-1}$]& 1.84$\pm$ 0.14 &  1.84$\pm$ 0.14 \\ 
  \hline
Nominal period and error [day]&4.3114$\pm$0.0002&4.3114$\pm$0.0002\\
\hline
Phase at BJD$_{ref}$ (epoch 2018)&0.6108$\pm$ 0.0086&0.6266$\pm$ 0.0087\\
\hline
 Zero crossing time T$_2$ at epoch 2018 (bjd-2.4e6)&58471.866$\pm$0.037&58471.802$\pm$0.037\\ 
  \hline
  Elapsed time T$_2$-T$_1$ [day]&3909.1$\pm$0.09&3909.03$\pm$0.09\\
  \hline
  K$_{per}$ number of periods in T$_2$-T$_1$&906.688&906.673\\
  \hline
  K1&906&906\\
  \hline
  K2&907&907\\
  \hline
  Period and error for K1 [day]&4.3147$\pm$0.0001&4.3146$\pm$0.0001\\
  \hline
  Period and error for K2 [day]&4.3099$\pm$0.0001&4.3098$\pm$0.0001\\
  \hline
\end{tabular}
 \caption{New estimation of period for planet HD 40307 b}
    \label{tab:HD40307planetb}
\end{table*}

\begin{table*}[h!]
\begin{tabular}{ |c|c|c|} 
  \hline
&Planet c pipeline&Planet c CF1\\
\hline
   Amplitude and error [ms$^{-1}$]& 2.29$\pm$ 0.13 &  2.29$\pm$ 0.13 \\ 
  \hline
Nominal period and error [day]&9.6210$\pm$0.0008&9.6210$\pm$0.0008\\
\hline
Phase at BJD$_{ref}$ (epoch 2018)&0.90041$\pm$0.0145&0.82599$\pm$0.0055\\
\hline
 Zero crossing time T$_2$ at epoch 2018 (bjd-2.4e6)&58465.837$\pm$0.1396&58466.470$\pm$0.1396\\ 
  \hline
  Elapsed time T$_2$-T$_1$ [day]&3915.747$\pm$0.88&3914.94$\pm$0.159\\
  \hline
  K$_{per}$ number of periods in T$_2$-T$_1$&406.85&406.92\\
  \hline
  K1&406&406\\
  \hline
  K2&407&407\\
  \hline
  Period and error for K1 [day]&9.6411$\pm$0.0005&9.643$\pm$0.0004\\
  \hline
  Period and error for K2 [day]&9.6174$\pm$0.0005&9.619$\pm$0.0004\\
  \hline
\end{tabular}
 \caption{New estimation of period for planet HD 40307 c}
    \label{tab:HD40307planetc}
\end{table*}

\begin{table*}[h!]
\begin{tabular}{ |c|c|c|} 
  \hline
&Planet d pipeline&Planet d CF1\\
\hline
   Amplitude and error [ms$^{-1}$]& 2.31$\pm$ 0.14 &  2.31$\pm$ 0.14 \\ 
  \hline
Nominal period and error [day]&20.412$\pm$0.004&20.412$\pm$0.004\\
\hline
Phase at BJD$_{ref}$ (epoch 2018)&0.39676$\pm$0.0419&0.33147$\pm$0.0555\\
\hline
 Zero crossing time T$_2$ at epoch 2018 (bjd-2.4e6)&58466.4&58467.737\\ 
  \hline
  Elapsed time T$_2$-T$_1$ [day]&3933.98$\pm$1.15&3935.32$\pm$1.15\\
  \hline
  K$_{per}$ number of periods in T$_2$-T$_1$&192.7287&192.79\\
  \hline
  K1&192&192\\
  \hline
  K2&193&193\\
  \hline
  Period and error for K1 [day]&20.489$\pm$0.00455&20.4964$\pm$0.0059\\
  \hline
  Period and error for K2 [day]&20.383$\pm$0.00457&20.3902$\pm$0.0059\\
  \hline
\end{tabular}
\caption{New estimation of period for planet HD 40307d}
    \label{tab:HD40307planetd}
\end{table*}

\end{appendix}

\end{document}